\documentclass{aa}
\usepackage[english]{babel}
\usepackage[varg]{txfonts}
\usepackage[usenames,dvipsnames]{xcolor}

\usepackage{epsfig}
\usepackage{gensymb}
\usepackage{graphicx,natbib}
\usepackage{amsmath,amssymb}
\usepackage{hyperref}
\usepackage{natbib}
\usepackage{tabularx}
\usepackage{makecell}

\usepackage{tikz}
\usetikzlibrary{arrows,shapes,positioning}
\tikzstyle{arrow}=[
    thick,
    ->,
    >=latex
    ]
\tikzstyle{nodebox}=[
    draw,
    fill=gray!15
    ]

\hypersetup{
    colorlinks=true,
    citecolor=blue,
    linkcolor=red,
    urlcolor=black
    }  

\nolinenumbers

\begin{document}

\def\nat{Nature}
\def\apj{ApJ}
\def\apjs{ApJS} 
\def\apjl{ApJ}
\def\apss{Ap. Sp. Sci.}
\def\aap{A\&A}
\def\mnras{MNRAS}
\def\jgr{JGR}
\def\zat{Zeitschrift f\"ur Astrophysik}

\def\cs{c_\mathrm{s}}
\def\kb{k_\mathrm{B}}
\def\mp{m_\mathrm{p}}
\def\lc{l_\mathrm{c}}
\def\Lc{L_\mathrm{c}}
\def\rc{r_\mathrm{c}}
\def\rd{r_\mathrm{disk}}
\def\tm{T_\mathrm{m}}
\def\ri{r_\mathrm{i}}
\def\md{\mathcal{M}_\mathrm{d}}
\def\Omk{\Omega_\mathrm{K}}
\def\rhoc{\rho_\mathrm{c}}
\def\fenh{f_\mathrm{enh}}
\def\nh{n_\mathrm{H}}
\def\tff{t_\mathrm{ff}}
\def\arcsec{\hbox{$^{\prime\prime}$}}
\def\kms{km s$^{-1}$}

\def\tempo#1{\noindent{\color{pink}#1}}
\def\newtxt#1{\noindent{\textbf{#1}}}

\title{APE: An analytical protostellar environment to provide physical conditions to chemical models and synthetic observations}
\titlerunning{APE: Analytical Protostellar Environment}

\author{P. Marchand\inst{1}, A. Coutens\inst{1}, A. Espagnet\inst{1}, F. Cruz-S{\'a}enz de Miera\inst{1,2}, J.-C. Loison\inst{3}, V. Wakelam\inst{4}}
\institute{Institut de Recherche en Astrophysique et Planétologie, Université de Toulouse, CNRS, CNES, 9 av. du Colonel Roche, 31028 Toulouse Cedex 4, France \\ \email{pierre.marchand.astr@gmail.com}
\and Konkoly Observatory, HUN-REN Research Centre for Astronomy and Earth Sciences, MTA Centre of Excellence, Konkoly-Thege Miklós út 15–17, 1121 Budapest, Hungary
\and Institut des Sciences Mol\'eculaires (ISM), CNRS, Univ. Bordeaux, 351 cours de la Lib\'eration, F-33400 Talence, France
\and Laboratoire d’astrophysique de Bordeaux, Univ. Bordeaux, CNRS, B18N, allée Geoffroy Saint-Hilaire, 33615 Pessac, France}

\authorrunning{P. Marchand}

\date{}

\abstract{Chemical modeling and synthetic observations are powerful methods to interpret observations, both requiring a knowledge of the physical conditions. In this paper, we present the Analytical Protostellar Environment (APE) code, which aims at making chemical simulations and synthetic observations accessible. APE contains a physical model of protostellar evolution (including the central object, the envelope, the protoplanetary disk and the outflow) as well as interfaces to publicly available codes to perform chemical simulations, radiative transfer calculations, and synthetic interferometry imaging. APE produces density and temperature maps of protostellar systems. The code can also follow individual particles throughout their journey in a collapsing core. APE includes a treatment of the dust grain size-distribution to compute opacities self-consistently for subsequent radiative transfer. We show an example of application of APE by computing chemical abundance maps of CO, CN, CS, H$_2$CO, and CH$_3$OH in a Class I protostellar system. We also performed synthetic ALMA observations of their molecular emission assuming an edge-on source inclination. The moment 0 maps of CO, CS, and H$_2$CO display an X-shaped emission similar to what is observed toward the Class I source IRAS~04302+2247.}


\maketitle

\section{Introduction}

Protostars harbor a rich and complex chemistry, both in their envelope and their protoplanetary disk \citep[e.g.,][]{Jorgensen2016,McGuire2022}. Protostars are however chemically diverse \citep{Lefloch2018}, highlighting the importance of understanding the chemical and physical processes that drive this diversity.

Interferometers such as the Atacama Large Millimeter/submillimeter Array (ALMA) and the NOrthern Extended Millimeter Array (NOEMA), with their high angular resolution ($<1 \arcsec$), can probe the chemical content of protostellar sources at the scale of protoplanetary disks. Coupled with spectrometers, those instruments deliver observations in the form of hyperspectral cubes. Although rigorous analysis may distinguish the origin of molecular emission \citep[the disk or the outflow for example, see e.g.,][]{Tychoniec2021}, interferometric observations remain complex to analyze.
Chemical models are a complementary tool to understand the origin of molecules in such regions. By carrying out radiative transfer calculations on simulated chemical abundance maps, we can produce synthetic (i.e., simulated) observations \citep[as in][]{Nazari2022}, which allows for a more direct comparison between models and interferometric observations.

Such models require physical conditions, typically density and temperature, as input for their calculations. They are usually provided by analytical physical models \citep[e.g.,][]{Visser2009,Gavino2021,Murillo2022} or by magnetohydrodynamics (MHD) simulations \citep[e.g.,][]{Coutens2020,Navarro2024}. MHD simulations however require large amounts of resources. Simulating the collapse of a prestellar core, followed by the formation and evolution of a central object and a protoplanetary disk, is extremely challenging due to the wide range of spatial and temporal scales needed to describe those phenomena \citep{Vaytet2018,Ahmad2024}. Analytical and semi-analytical models require much less computational resources, allowing for broader accessibility and making parametric studies affordable. These models rely on simplifying assumptions and cannot fully capture the complexity of the underlying physics. On the other hand, they provide clear relationships between key physical parameters and their macroscopic results (e.g., the impact of the magnetic mass-to-flux ratio on the size of protoplanetary disks), which greatly facilitates the interpretation.

In this paper, we present the Analytical Protostellar Environment (APE) code\footnote{\url{https://bitbucket.org/pmarchan/ape_code}}. APE simulates density, temperature, extinction, and dust-size distribution maps in a protostellar environment, including the central object, the envelope, the disk, and the outflow. The analytical model can simulate the whole system evolution throughout the prestellar core, Class 0 and Class I phases, at a fraction of the cost of MHD simulations. In addition to the physical model, we provide scripts and interfaces to other publicly available codes to perform chemical simulations \citep[Nautilus,][]{Ruaud2016}, radiative transfer calculations \citep[RADMC-3D,][]{Dullemond2012}, and synthetic imaging (Imager\footnote{\url{http://www.iram.fr/IRAMFR/GILDAS}\\ \url{https://imager.oasu.u-bordeaux.fr}}). The aim of APE is to provide a fast and accessible way to generate maps of chemical abundances and their simulated emission. 

Section \ref{sec:ape} is dedicated to the description of APE, with Sect. \ref{sec:ape_general} explaining in more detail the general purpose of the code. In Sect. \ref{sec:application}, we show an example of application. We conclude in Sect. \ref{sec:conclusion}.

\section{The APE code}\label{sec:ape}

\subsection{General description}\label{sec:ape_general}

\begin{figure} 
    \centering
    \includegraphics[width=0.5\textwidth, trim=0cm 0cm 0.1cm 0.1cm,clip]{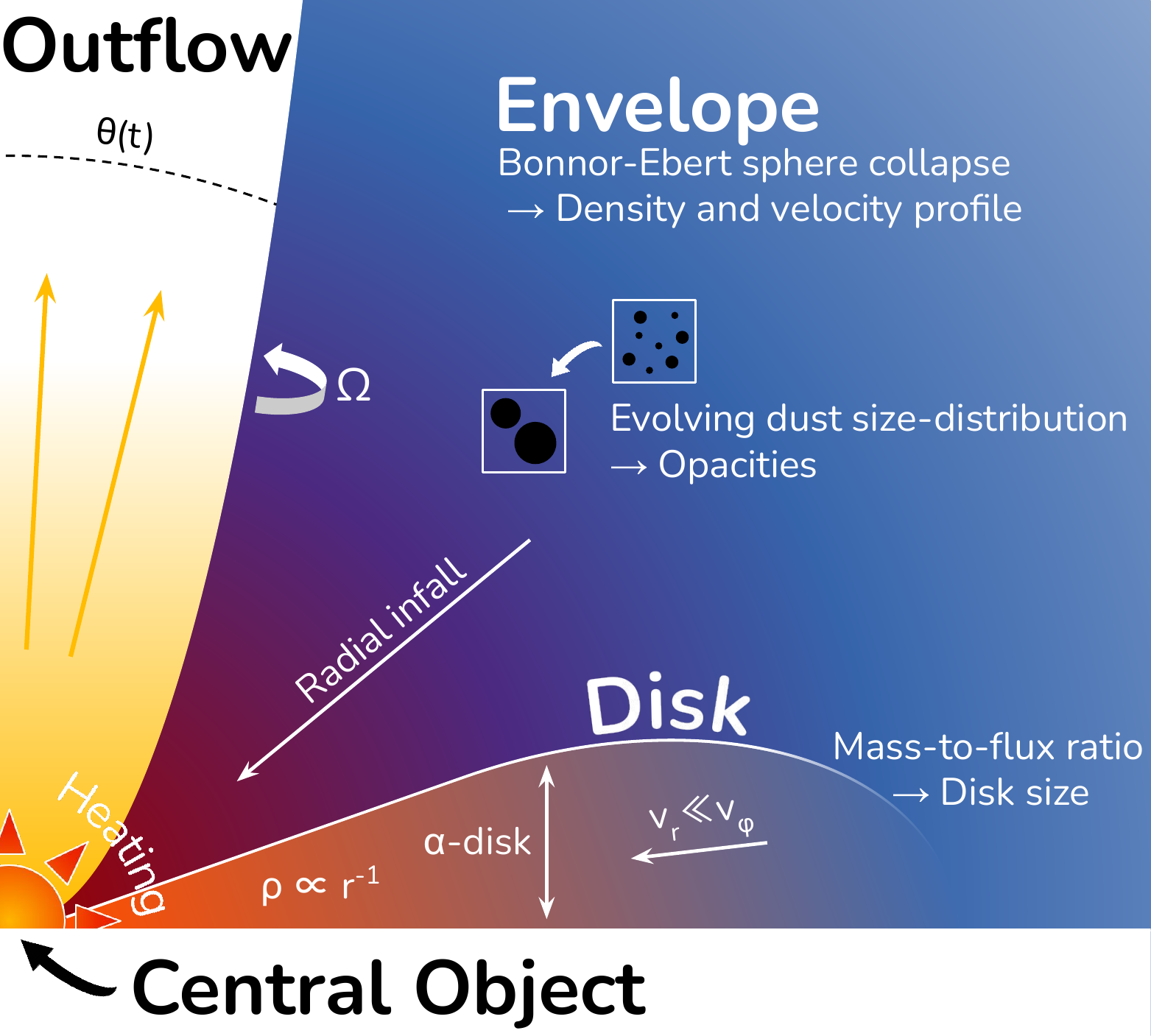}
    \caption{Schematic of an APE snapshot simulation.}
     \label{fig:ape_summary}
\end{figure}

APE can be used in two modes: a “snapshot” mode, in which it generates a map of density and temperature, and a “particle” mode, which calculates the trajectory of a virtual particle in the evolving system.
The initial condition ($t=0$) is a critical Bonnor-Ebert sphere \citep{Ebert1955,Bonnor1956} for which we derived analytical evolution equations for its dynamics and density profile. After a free fall time, a central object, a disk and an outflow are created. The mass accreted in the center is distributed between the central object and the disk, from which the code computes the radius of the central object, luminosity and temperature, and the disk density and temperature profiles. Dust grains are included in the form of an evolving size-distribution. A schematic of the simulated region is displayed in Fig. \ref{fig:ape_summary}. All APE input parameters are listed in Table \ref{tab:ape_parameters} with their description.

\subsubsection{Working modes}

\paragraph{Snapshot mode}

\begin{figure} 
    \centering
    \includegraphics[width=0.5\textwidth, trim=0cm 0cm 0cm 0cm]{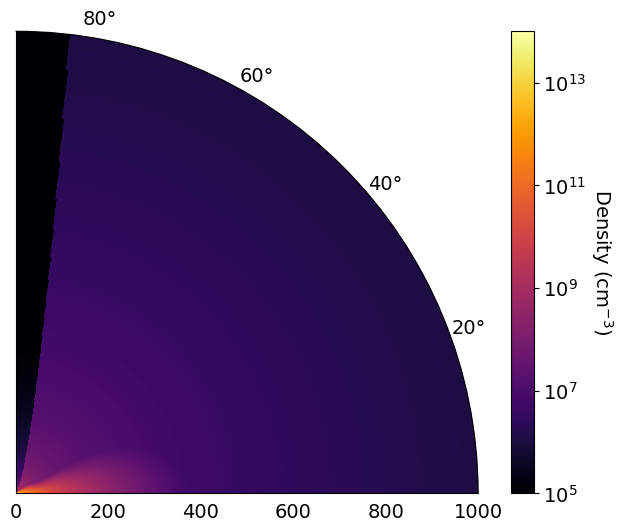}
    \includegraphics[width=0.5\textwidth, trim=0cm 0cm 0cm 0cm]{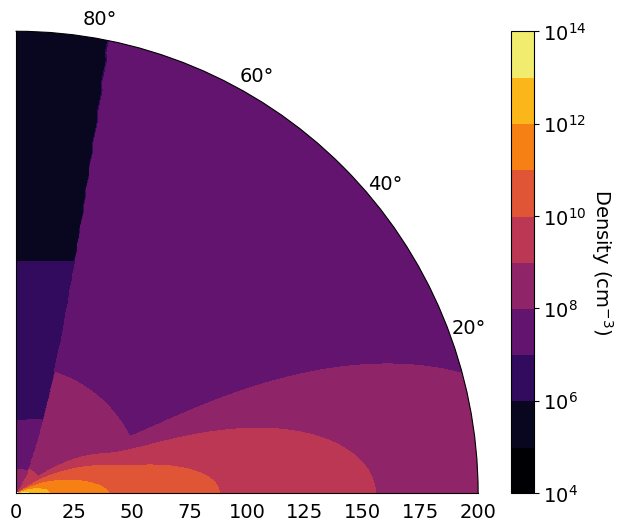}
    \caption{Top panel: Density map generated by APE. Bottom panel: Zoom of the inner 200 au, with a reduced color sampling.}
     \label{fig:ape_map}
\end{figure}

In snapshot mode, APE generates a map of a protostellar system with an envelope, a disk, a central object and an outflow. The code uses a 2D grid in spherical coordinates, assuming a cylindrical symmetry and a symmetry with respect to the mid-plane of the disk. The region simulated, in spherical coordinates, is then [$0\leq r \leq l_\mathrm{box}$, $\varphi=0$, $0 \leq \theta \leq \pi/2$], with $l_\mathrm{box}$ the size of the simulated region (input parameter). The user can either choose a linear or a logarithmic grid spacing in radius. In each cell, the flow variables (density, temperature, velocities and the grain size-distribution) are calculated as a function of the model parameters input by the user: the mass of the initial Bonnor-Ebert sphere, the magnetic mass-to-flux ratio and the age of the central object. An example of a snapshot is displayed in Fig. \ref{fig:ape_map}. Generating the full map takes typically 1 second for 10 000 grid cells on a regular laptop.

\paragraph{Particle mode}

The purpose of the particle mode is to emulate hydrodynamics simulations at a fraction of the cost. The user chooses an initial position of the particle at $t=0$ (in the initial Bonnor-Ebert sphere). The trajectory of the particle is then calculated over time, until the particle reaches the center (the central object if it is created), the outflow or until a time defined by the user. The result is a complete density and temperature history of the particle.
This computation can be done in reverse, with the user choosing the age of the system, like in snapshot mode, and the final position of the particle at this time. The trajectory is then calculated backward until the system reaches its initial condition at $t=0$. Five to thirty seconds are needed to compute the full trajectory.

\paragraph{Grid of particles}

The snapshot and particle modes can be combined to compute the full history of the gas displayed in a snapshot.
In that case, a map is created like in snapshot mode, and a particle placed in each grid cell. APE then computes the trajectory in reverse of all those particles until $t=0$. The density and temperature history of those particles can then be used as an input for a chemical modeling code, to ultimately produce maps of chemical abundances.

\subsubsection{Radiative transfer}\label{sec:radiative_transfer}

APE has been designed to work with the radiative transfer code RADMC-3D \citep{Dullemond2012}. In snapshot mode, the output files of APE can be directly used as input for RADMC-3D with opacities calculated directly from the dust grain model of APE. That way, the user can either generate a temperature map more accurate than the analytical descriptions provided in APE, or address any science topic for which radiative transfer calculations are needed.   
APE also offers the possibility to use temperatures from RADMC-3D simulations in particle mode. Since running radiative transfer calculations at each time-step would be costly, we provide scripts to generate snapshots at regular time intervals, on which RADMC-3D is run to produce temperature maps at those times. Subsequent APE simulations in particle mode then use those maps to interpolate (in time and space) the temperature at every time-step in the particle trajectory.

\begin{table*}
  \caption{APE input parameters}
  \label{tab:ape_parameters}
\centering
\begin{tabular}{lllll}
\hline\hline
  Name & Unit & Description & Main impact on the model & Recommended range \\
\hline
  & & \hspace{3cm} Grid parameters & & \\
\hline
  loggrid   & Boolean & Logarithmic grid or linear & Spatial resolution & .true. -- .false. \\
  rad\_min\_au   & au & Innermost cell radius & Spatial resolution & 0.1 -- 100 \\
  rad\_max\_au   & au & Outermost cell radius & Simulated region size & 50 -- 10 000 \\
  nrad   & Integer & Number of radial cells & Spatial resolution & 10 -- 1000 \\
  theta\_min\_deg   & degrees & Minimum polar angle & Simulated region size & 0 -- 89 \\
  theta\_max\_deg   & degrees & Maximum polar angle & Simulated region size & 1 -- 90 \\
  ntheta   & Integer & Number of polar cells & Polar resolution & 10 -- 100 \\
  radmc\_output   & Boolean & Produce RADMC-3D-ready inputs & None & .true. -- .false. \\
  snapshot\_nautilus & Boolean & Produce Nautilus-ready inputs on a snapshot & None & .true. -- .false. \\
  nautilus\_time\_kyr & kyr & Duration of the nautilus simulation & None & 0 -- 20 \\
\hline
  & & \hspace{3cm} Model parameters & & \\
\hline
  Mass  & M$_\odot$ & Mass of the Bonnor-Ebert sphere & Envelope density, $\tff$ & 0.2 -- 8 \\
  time\_years   & years & Time after the start of the collapse & Envelope, star and disk & 0 -- $2 \tff~(\sim 3\times 10^5$) \\
  t\_pstar\_age   & Boolean & Is time\_years the age of the central object $^a$ & Same as time\_years & .true. -- .false. \\
  temp\_mol\_cloud   & K & Temperature of the surround molecular cloud & Temperature in the system & 5 -- 30 \\
  masstoflux   & None & Mass to flux ratio ($\propto 1/B$)  & Disk size & 1 -- $\infty$ \\
  coagulation  & Boolean & Enable estimation of dust growth & Radiative transfer & .true. -- .false. \\
  dust\_to\_gas  & None & Desired dust-to-gas mass ratio & Radiative transfer & 0.001 -- 0.1 \\
  alphadisk  & None & $\alpha$-disk parameter & Disk scale height & 0.0001 -- 0.1 \\
  disk\_cutoff  & Boolean & If .true., uses a sharp cut-off of the disk & Disk profile & .true. -- .false. \\
  use\_radmc\_temp & Boolean & If .true. use RADMC-3D temperature & Gas temperature & .true. -- .false.\\
\hline
  & & \hspace{3cm} Particle parameters & & \\
\hline
  x\_ini, z\_ini & au & Initial x (cylindrical radius) and z coordinates & Particle trajectory & 0 -- 20000 \\
  tmax & years & Maximum time for the particle evolution & None &  0 -- $\infty$\\
  dt0 & years & Maximum allowed time-step& Trajectory accuracy &  1 -- 1000\\
  dyn\_fact & None & Time-step limitation near box edges & Trajectory accuracy &  100 -- 1000\\
  Omega0   & rad s$^{-1}$ & Angular velocity of the cloud  & Envelope rotation & $10^{-15}$ -- $10^{-12}$ \\
  cloud\_extinction   & mag & Visual extinction of the surrounding cloud  & Chemical post-processing & 0 -- $\infty$ \\
  reverse & Boolean & Compute particle trajectories backward in time & None & .true -- .false.\\
\hline
\end{tabular}
\tablefoot{$^a$ By default, the parameter time\_years represents the time after the start of the collapse of the initial Bonnor-Ebert sphere. However, the user may find more convenient to choose directly a time after the formation of the central object. This can be done by adding $\tff$ to the desired age for the central object, but setting the parameter t\_pstar\_age to .true. does this automatically. In that case, the parameter time\_years then represents the time after the formation of the central object.}
\end{table*}

\subsection{The envelope}

The envelope initial condition is a critical Bonnor-Ebert sphere \citep{Ebert1955,Bonnor1956} in free fall.

\subsubsection{Base of the model}

\begin{figure}
    \centering
    \includegraphics[width=0.5\textwidth, trim=2.5cm 2cm 0cm 2cm]{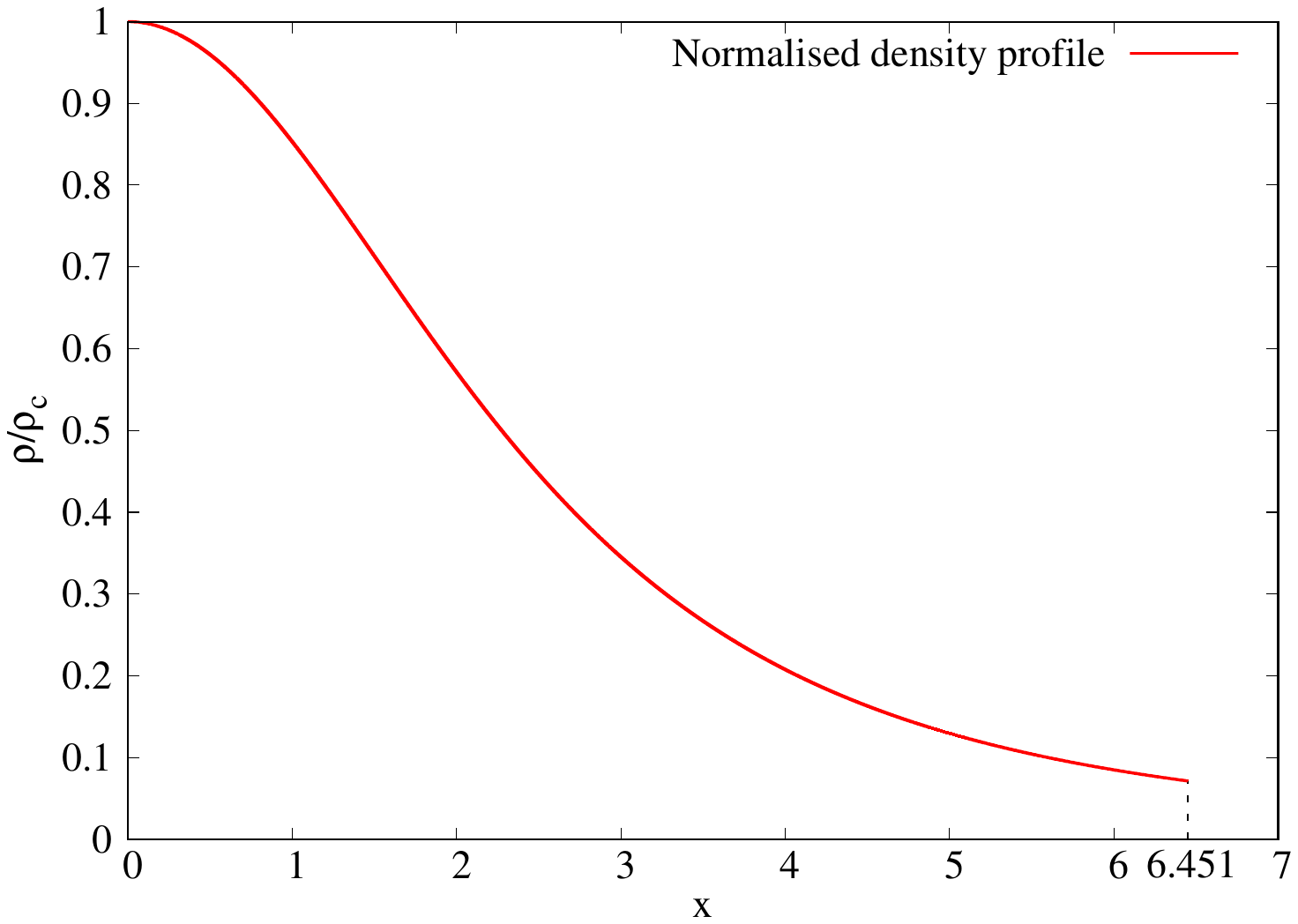}
    \caption{Normalized density profile of a Bonnor-Ebert sphere.}
    \label{fig:be}
\end{figure}

The Bonnor-Ebert sphere is a spherically symmetric isothermal cloud of gas in which the gravity and the thermal pressure are at equilibrium. The density profile, $\rho$, is normalized by the central density, $\rhoc$,
\begin{equation}
    D=\frac{\rho}{\rhoc},
\end{equation}
while the spherical radius coordinate, $r$, is normalized as
\begin{equation}\label{eq:be:norm}
  x_\mathrm{be} = r \frac{\sqrt{4\pi G \rhoc}}{\cs},
\end{equation}
with $G$ the gravitational constant and $c_\mathrm{s}$ the sound speed. The equilibrium between thermal pressure, $P$, and gravity potential, $\Phi$, reads
\begin{equation}\label{eq:be:therm_eq}
    \nabla{\Phi}=-\frac{1}{\rho}\nabla P.
\end{equation}
We can combine eq. \ref{eq:be:therm_eq} with the equation of state for an isothermal perfect gas
\begin{equation}
    P=\rho \cs^2,
\end{equation}
to obtain the relation
\begin{equation}\label{eq:be:rhophi}
  \rho = \rhoc e^{-\frac{\Phi}{\cs^2}}.
\end{equation}
Finally, combining the Poisson equation
\begin{equation}
    \Delta{\Phi}=4\pi G\rho,
\end{equation}
with the divergence of eq. \ref{eq:be:therm_eq} in spherical coordinates results in the Lane-Emden equation
\begin{equation} \label{eq:be:LE}
  \frac{\mathrm{d}^2\Phi}{\mathrm{d}x_\mathrm{be}^2} + \frac{2}{x_\mathrm{be}} \frac{\mathrm{d}\Phi}{\mathrm{d}x_\mathrm{be}} = e^{-\frac{\Phi}{\cs^2}},
\end{equation}
The normalized radius of a critical Bonnor-Ebert sphere is $x_\mathrm{be}=x_\mathrm{crit}\approx6.451$, above which no hydrostatic equilibrium is possible.
The gravitational potential and density profiles, $\Phi(x_\mathrm{be})$ and $\rho(x_\mathrm{be})$, are integrated using the Lane-Emden equation (\ref{eq:be:LE}), with $\Phi(0)=0$ and $\mathrm{d}\Phi/\mathrm{d}x_\mathrm{be} (0)=0$.
The resulting Bonnor-Ebert density profile, $D(x_\mathrm{be})$, is represented in Fig. \ref{fig:be}.

The mass of the cloud of radius $r_\mathrm{cld}$ is given by
\begin{align}\label{eq:be:mass}
  M_\mathrm{cld} = & \int_0^{r_\mathrm{cld}} 4\pi r^2 \rho(r) dr, \nonumber\\
    = & 4\pi \rhoc \left(\frac{\cs}{\sqrt{4\pi G \rhoc}}\right)^3 \int_0^{x_\mathrm{crit}} D(x) x_\mathrm{be}^2 dx_\mathrm{be}, \nonumber\\
    = & I_m \cs^3 G^{-\frac{3}{2}} (4\pi)^{-\frac{1}{2}} \rhoc^{-\frac{1}{2}}.
\end{align}
where $I_m$ is the value of the integral on $x$ that approximates to $15.7$ when integrating the whole profile. Finally the central density is given by
\begin{equation}\label{eq:be:rhoc}
  \rhoc= I_m^2 \cs^6 G^{-3} (4\pi)^{-1} M_\mathrm{cld}^{-2}= 7.78 \times 10^{-19} \mathrm{g~cm}^{-3} \left( \frac{M_\mathrm{cld}}{M_\odot}\right)^{-2}.
\end{equation}

We assumed that the enclosed mass is invariant with time for a given fluid particle (equivalent to collapsing spherical shells that cannot cross each-other). The free fall velocity of a particle at radius r would then be $v_r(r)=-\sqrt{2GM_\mathrm{in}(r)/r}$, where $M_\mathrm{in}(r)$ is the mass enclosed at radius r. However, several physical processes support the gas against gravity, effectively reducing the infall velocity: rotation, thermal pressure, magnetic fields and turbulence. The effective free fall time can then be increased by a factor $\sim 1.3 - 2$ with rotation and thermal pressure alone \citep[see e.g.,][]{Gabbasov2017}, and by a factor $\gtrsim 2$ between moderate and strong magnetic fields \citep{Masson2016}. We therefore assumed a factor 2 reduction in radial velocity and adopt
\begin{equation}\label{eq:be:vel}
    v_r(r) = -\sqrt{\frac{GM_\mathrm{in}(r)}{2r}}.
\end{equation}
Solving the differential equation
\begin{equation}
    \frac{dr}{dt}=v_r(r),
\end{equation}
yields the following analytical expression
\begin{equation}\label{eq:be:rdet}
    r(t) = \left[ r_0^{\frac{3}{2}} - \frac{3}{2}\sqrt{\frac{G M_\mathrm{in}(r_0)}{2}}t \right]^{\frac{2}{3}}.
\end{equation}
where $r_0$ is the radius of the fluid particle at $t=0$.

\subsubsection{Computing the flow variables}\label{sec:envelope_density}

\begin{figure}
    \centering
    \includegraphics[width=0.5\textwidth, trim=0cm 0cm 0cm 0cm]{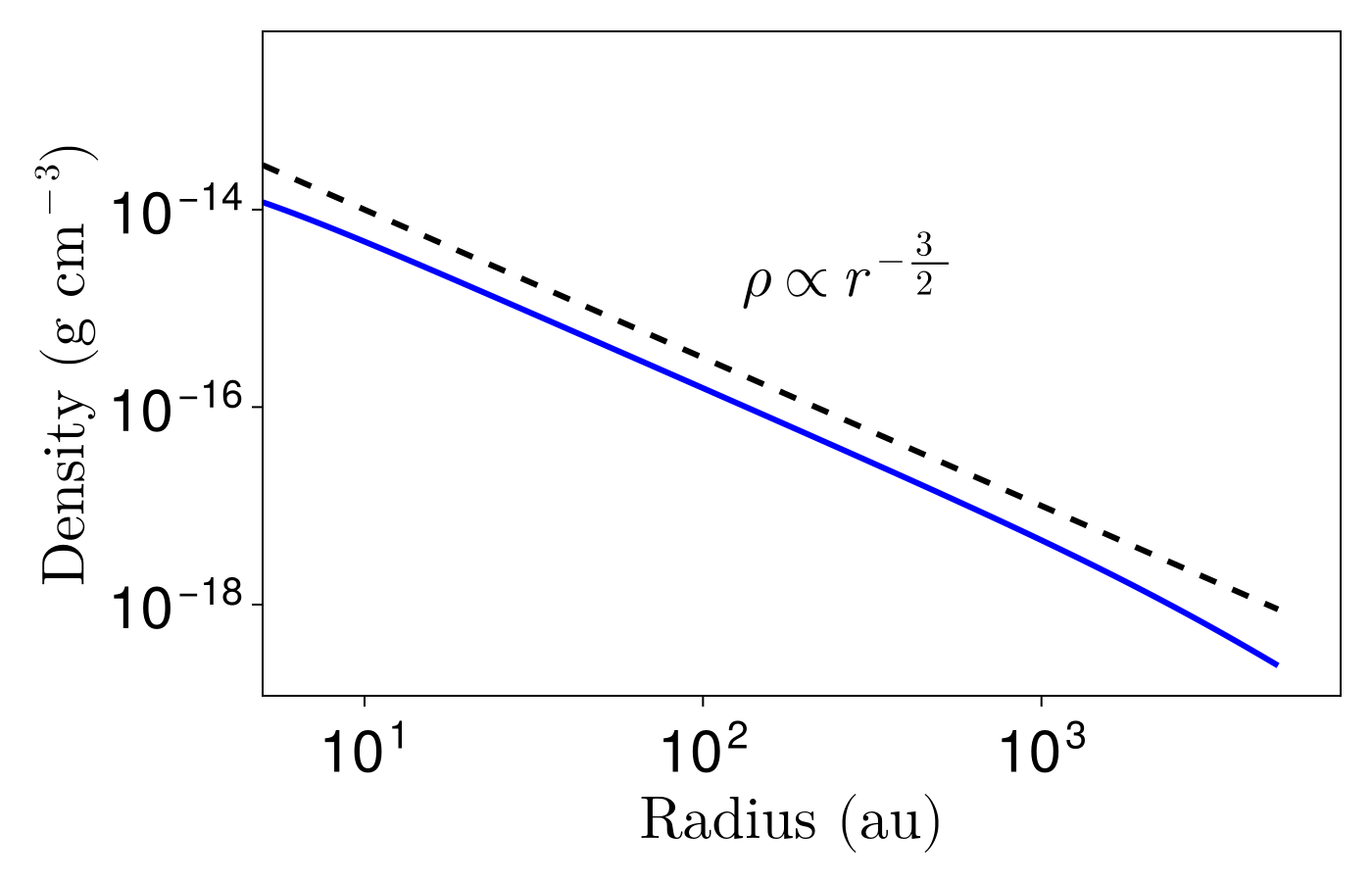}
    \caption{Density profile of a collapsing Bonnor-Ebert sphere following our model, for a 2 M$_\odot$ sphere at time $t\approx 192$ kyr (in blue). The dashed black line represents a $r^{-\frac{3}{2}}$ profile.}
    \label{fig:be:profile}
\end{figure}

The aim is to compute the density and velocity at any given time, $t$, and radius, $r$, in the envelope. Rewriting eq. (\ref{eq:be:rdet}) at a time $t$ and radius $r$, the equation
\begin{equation}\label{eq:be:r0}
    \frac{1}{r_0}\left[ r_0^{\frac{3}{2}} - \frac{3}{2}\sqrt{\frac{G M_\mathrm{in}(r_0)}{2}}t \right]^{\frac{2}{3}} - \frac{r}{r_0} = 0,
\end{equation}
is solved for $r_0(r,t)$ by dichotomy. With the initial position of the fluid particle, we obtain the enclosed mass by integrating eq. (\ref{eq:be:mass}), and the radial velocity with eq. (\ref{eq:be:vel}).
To calculate the density, we considered the compression of a thin spherical shell contained between radii $r$ and $r+\mathrm{d}r$. The density is then computed using the initial density of the fluid particle $\rho(r_0(r,t),t=0)$, from the initial Bonnor-Ebert density profile, and the volume compression of the spherical shell between $[r_0(r,t);r_0(r+dr,t)]$ and $[r,r+dr]$. We therefore have 
\begin{equation}\label{eq:be:rho}
    \rho(r,t) = \rho(r_0(r,t),t=0)\frac{r_0(r+\mathrm{d}r,t)^3 - r_0(r,t)^3}{(r+\mathrm{d}r)^3-r^3}.
\end{equation}
In the code, we take $\mathrm{d}r=r/100$. While the Bonnor-Ebert density profile is asymptotically proportional to $r^{-2}$, the slope transitions toward $r^{-1.5}$ as the collapse proceeds, similarly to other free fall analytical models \citep[e.g.,][]{Cassen1981,Hartmann2009}. Figure \ref{fig:be:profile} shows the density profile of the envelope for a 2 M$_\odot$ cloud at time $t\approx192$ kyr (free fall time of 152 kyr). To validate our envelope model, we also present a comparison between the envelope model of APE and a MHD simulation in Appendix \ref{app:envelope}.

The envelope rotation velocity has no impact on the physical model, but can be relevant for synthetic observations. We assumed that the envelope is initially rotating at an angular velocity of $\Omega_0$, which is chosen by the user. The angular velocity at a radius $r$ and time $t$ is then simply calculated assuming angular momentum conservation
\begin{equation}
  \Omega(r,t) = \Omega_0 \left(\frac{r_0(r,t)}{r}\right)^2.
\end{equation}

\subsubsection{Maximum time}
Assuming the cloud has a normalized radius of $x=x_\mathrm{crit}=6.451$, we can compute the cloud radius $r_\mathrm{cld}$ with eq. (\ref{eq:be:norm}). The edge of the cloud reaches the center at time $t_\mathrm{max}$, with
\begin{equation}
    r(t_\mathrm{max}) = \left[ r_\mathrm{cld}^{\frac{3}{2}} - \frac{3}{2}\sqrt{\frac{G M_\mathrm{cld}}{2}}t_\mathrm{max} \right]^{\frac{2}{3}} = 0.
\end{equation}
Therefore
\begin{equation}
    t_\mathrm{max} = \frac{2\sqrt{2}}{3\sqrt{GM_\mathrm{cld}}}r_\mathrm{cld}^{\frac{3}{2}}.
\end{equation}
Replacing $r_\mathrm{cld}$ with equation (\ref{eq:be:norm}), $M_\mathrm{cld}$ with eq. (\ref{eq:be:mass}), and $\rho_\mathrm{c}$ using the free fall time expression
\begin{equation}
    \tff = \sqrt{\frac{3\pi}{32G\rho_\mathrm{c}}},
\end{equation}
we obtain
\begin{equation}
    t_\mathrm{max} = \frac{8 x_\mathrm{crit}^{\frac{3}{2}}}{3\sqrt{3}\pi\sqrt{I_m}} \tff  \approx 2 \tff.
\end{equation}
Beyond 2 free fall time, we therefore considered by default that the envelope is depleted. Should the user wish to extend the envelope lifetime beyond 2 free fall times, APE provides the possibility to do so by changing an “effective” $x_\mathrm{crit}$ which only affects the extent of the initial Bonnor-Ebert sphere (not the central density or the free fall time).

\subsubsection{Envelope temperature}

In APE, the temperature may be computed using the radiative transfer code RADMC-3D \citep{Dullemond2012}, which provides a temperature at each location for each grain size. We also provide a size-independent approximation of the temperature for faster calculations, which we describe hereafter.
The initial envelope temperature $T_\mathrm{mc}$ is set by the user with the temp\_molecular\_cloud parameter. Once the central object is ignited and starts radiating from its surface, the temperature rises. Assuming the envelope is optically thin, the total energy flux is conserved when integrated over a sphere. Therefore, writing $\phi (r)$ the radiative power received per surface unit at radius $r$, the quantity $\phi (r) 4 \pi r^2$ is constant with respect to r. Assuming the central object is a black body, the radiative flux emitted per surface unit at the central object surface is
\begin{equation}\label{eq:temp:flux_star}
\phi_*(R_*) = \sigma_\mathrm{B} T_*^4,
\end{equation}
with $R_*$ and $T_*$ the radius and temperature of the central object, and $\sigma_\mathrm{B}$ the Stefan-Boltzmann constant.
The total radiative flux crossing a sphere of radius $r$ is then equal to the flux emitted by the central object
\begin{equation}
4\pi r^2 \phi(r) = 4\pi R_*^2 \phi_*(R_*).
\end{equation}
Assuming a background radiation from the surrounding molecular cloud at temperature T$_\mathrm{mc}$, a dust grain at a distance $r$ then receives the radiative power per surface unit
\begin{equation}\label{eq:temp:flux_in}
\phi_\mathrm{grain}(r) = \phi(r)+\sigma_\mathrm{B} T_\mathrm{mc}^4,
\end{equation}
At thermal equilibrium, assuming that the grains are black bodies, the received and emitted flux are equal.
\begin{equation}\label{eq:temp:flux_out}
\phi_\mathrm{grain}(r) = \sigma_\mathrm{B} T_\mathrm{grain}(r)^4.
\end{equation}
Combining eq. (\ref{eq:temp:flux_star}), (\ref{eq:temp:flux_in}) and (\ref{eq:temp:flux_out}), we obtain the grain equilibrium temperature independent of the grain size.
\begin{equation}\label{eq:temp_env}
T_\mathrm{grain}(r) = \left[ \left(\frac{R_*}{r}\right)^{2} T_*^4 + T_\mathrm{mc}^4 \right]^{\frac{1}{4}}.
\end{equation}
Assuming the gas and grains reach thermal equilibrium quickly, we write $T(r) \approx T_\mathrm{grain}(r)$. This model does not account for the shadow cast by the disk, which lowers the temperature of the envelope near the midplane. However, this temperature is simply an estimate that does not affect the density or dynamics of the envelope.
This relation provides a good approximation to the envelope temperature calculated by the radiative transfer computations of RADMC-3D, for $r\gtrsim 50$ au. Closer to the central object, the envelope temperature is higher along the walls of the outflow cavity. For a precise temperature map, we encourage users to perform the radiative transfer calculations with the tools provided with APE.

\subsection{The central object}

The typical density at which the first hydrostatic core collapses to form the central object is typically $10^{-8}$ g cm$^{-3}$ \citep{Larson1969,Vaytet2018}. This density is reached at around $t=\tff$ after the start of the collapse. We therefore set the creation of the central object and the disk at $t=\tff$.

\subsubsection{Mass and radius}

We defined the total accreted mass $M_\mathrm{acc}$ as the mass of the envelope that has collapsed within the innermost cells (typically on a scale of 1 au). This mass is distributed between the disk $M_\mathrm{disk}$ and the central object $M_*$. \citet{Adam1986} derive
\begin{equation} \label{eq:disk_mass}
  M_\mathrm{disk} = (1-\eta)\md M_\mathrm{acc},
\end{equation}
with $\eta$ a free parameter, and
\begin{equation}
  \md = \frac{1}{3} u_* \int_{u_*}^1 (1-u)^{\frac{1}{2}}u^{-\frac{4}{3}}du,
\end{equation}
where $u_*=R_*/\rd$. Following \citet{Young2005}, we take $\eta=0.75$ so that the ratio $M_\mathrm{disk}/M_*$ remains close to the observationally estimated value of $\approx 1/3$ \citep[][]{Li2002,Kratter2008}. The star mass is therefore
\begin{equation} \label{eq:protostar_mass}
    M_* = M-M_\mathrm{disk} = (1-(1-\eta)\md)M_\mathrm{acc}.
\end{equation}

\begin{figure}
    \centering
    \includegraphics[width=0.5\textwidth, trim=0cm 0cm 0cm 0cm]{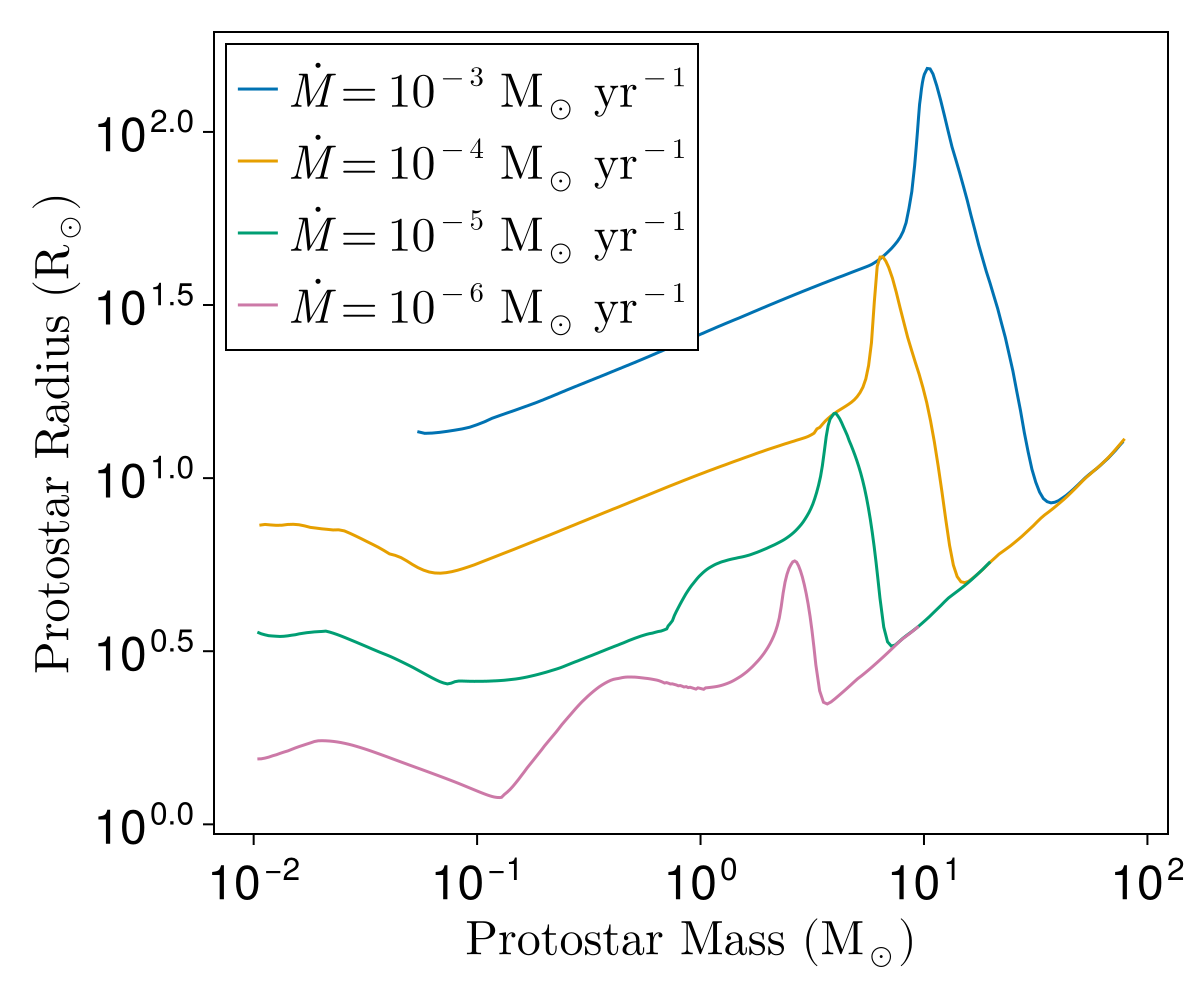}
    \includegraphics[width=0.5\textwidth, trim=0cm 0cm 0cm 0cm]{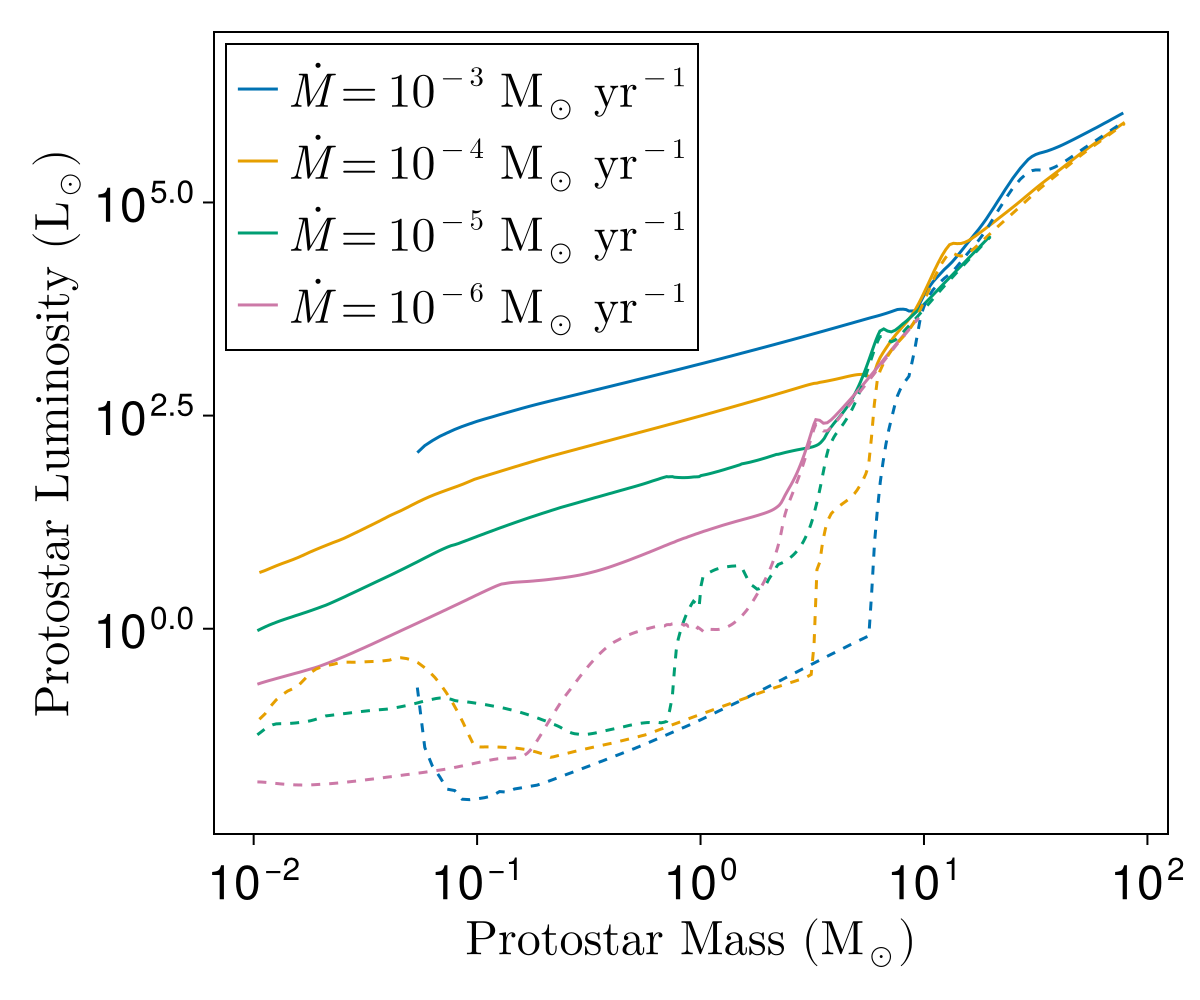}
    \caption{Upper panel: Mass-radius relation of central objects for different accretion rates, from the tabulated results of \citet{Hosokawa2009}. Bottom panel: Same as upper panel for the mass-luminosity relation. The solid lines represent the total central object luminosity $L_*$, while dashed lines represent only the photometric luminosity (excluding the accretion luminosity).}
    \label{fig:ho09_tracks}
\end{figure}

The radius of the central object, $R_*$, is controlled by the increase in entropy at the accretion shock on its surface, as well as the extent of its convection zone and the deuterium-burning zone. 
This calculation was performed in detail by \citet{Hosokawa2009} for low- and high-mass central objects. We used tabulated results from their simulations that provide the evolution of the radius and luminosity of young stars as a function of their mass and accretion rate. The instantaneous accretion rate $\dot{M}$ is calculated at any time $t$ as the variation in the accreted mass $M_\mathrm{acc}$ between times $0.99t$ and $1.01t$. We assumed that the central object receives 75\% of this mass to match the approximate mass ratio between the disk and the central object. We then paired the evolution tracks of \citet{Hosokawa2009} with equation (\ref{eq:protostar_mass}) and determine the corresponding $M_*$ and $R_*$ by dichotomy and interpolation of the tables. The upper panel of Fig. \ref{fig:ho09_tracks} shows the mass-radius relation for several accretion rates. The radius generally increases with mass, although central objects undergo a phase of swelling and contraction above 3 M$_\odot$. Those relations have been computed up to 10 M$_\odot$ for $\dot{M} = 10^{-6}$ M$_\odot$ yr$^{-1}$ and 90 M$_\odot$ for $\dot{M} = 10^{-3}$ M$_\odot$ yr$^{-1}$.

\subsubsection{Luminosity and temperature}

We interpolated the central object luminosity $L_*$ from the tabulated results of \citet{Hosokawa2009}, using the calculated accretion rate and mass of the central object. $L_*$ includes both the photometric luminosity and the accretion luminosity. The bottom panel of Fig. \ref{fig:ho09_tracks} displays both the photometric and total luminosities as a function of $M_*$ for different accretion rates. 
The effective temperature of central object is then calculated from $L_*$ and $R_*$ as
\begin{equation} \label{eq:protostar_temperature}
    T_* = \left( \frac{L_*}{4\pi \sigma_\mathrm{B} R_*^2} \right)^{\frac{1}{4}}.
\end{equation}

\subsection{The disk}

\subsubsection{The magnetic field and disk radius}

Magnetic fields are important for the formation and evolution of protoplanetary disks. The field is coupled to the gas and gets twisted by its rotation. That creates a magnetic tension force that opposes the rotation, which is called magnetic braking. In an ideal magnetohydrodynamics (MHD) description, the magnetic field is perfectly coupled to the gas. The field piles-up in the central region and creates a strong braking, preventing the formation of a rotationally supported disk (RSD). The non-ideal MHD description, which accounts for dissipative and decoupling processes of the magnetic field, is needed to prevent such a strong braking and the unbounded pile-up of the magnetic field.

\citet{Lee2021} derived analytical formulae for the radius of RSDs under a full non-ideal magnetohydrodynamics (MHD) regime. They assume that at equilibrium, the accretion of angular momentum compensates exactly the loss by magnetic braking, and consider relevant non-ideal MHD effects in different regimes. In the present model, we do not explicitly calculate the magnetic field geometry, but we can estimate the strength of the magnetic field in the disk. As highlighted by \citet{Masson2016} and \citet{Vaytet2018}, the magnetic field strength reaches a plateau in the disk region due to its dissipation by diffusive effects. In the envelope, the magnetic field intensity increases with the contraction of the gas. Following \citet{Nakano2002}, we have
\begin{equation}\label{eq:bfield}
  B=\frac{1}{\lambda} 4\left( \frac{\pi \kb T \rho}{\mu \mp} \right)^{\frac{1}{2}} = 4.27 \times 10^{-2}~\mathrm{G} \left( \frac{\rho}{10^{-13}~\mathrm{g}~\mathrm{cm}^{-3}}\right)^{\frac{1}{2}} \lambda^{-1},
\end{equation}
where $\kb$ is the Boltzmann constant, $\mu=2.31$, the mean atomic mass of molecules in the gas, and $\mp$ the mass of the proton. $\lambda$ is the mass-to-flux ratio, which is defined by \citet{Mouschovias1976} as
\begin{equation}
  \lambda = \frac{M/\Phi_\mathrm{B}}{\left(M/\Phi_\mathrm{B}\right)_\mathrm{crit}},
\end{equation}
\noindent with $M$ the cloud mass, $\Phi_\mathrm{B}=\pi r_\mathrm{cld}^2 B$ the magnetic flux and
\begin{equation}
  \left(\frac{M}{\Phi_\mathrm{B}}\right)_\mathrm{crit}=\frac{0.53}{3\pi}\sqrt{\frac{5}{G}}.
\end{equation}
We therefore have $\lambda \propto B^{-1}$. The mass-to-flux ratio is the main parameter influencing the disk size, and can take any value $>1$ \citep{Crutcher1999}. Numerical simulations generally consider $\lambda$ values between 2 and 10 for magnetized cores. We calculated the magnetic field strength at $\rho=10^{-13}$ g cm$^{-3}$, which is the formation density of the first hydrostatic core \citep{Larson1969}, which coincides with the formation of the disk.

\citet{Lee2021} determine two possible outcomes that are the “para-disk” and the “ortho-disk” \citep[see also][]{Tsukamoto2015}, depending on the relative directions of the disk rotation and the Hall drift of magnetic field lines. Ultimately, they show that the ortho-disk is short-lived and that a counter-rotating para-disk eventually forms, which is also observed in simulations \citep{Tsukamoto2017,Zhao2020}. \citet{Hennebelle2016} and \citet{Lee2021} derive the radius of the para-disk

\begin{equation} \label{eq:paradisk}
  \rd(t) = 19.2~\mathrm{au}~ \left[ \frac{\eta_\mathrm{A}}{10^{19} \mathrm{cm}^2~\mathrm{s}^{-1}}\right]^{\frac{2}{9}} \left[ \frac{M_\mathrm{acc} (t)}{0.1 \mathrm{M}_\odot}\right]^{\frac{1}{3}} \left[ \frac{B}{0.1 \mathrm{G}}\right]^{-\frac{4}{9}}.
\end{equation}
Using our magnetic field prescription, eq. (\ref{eq:paradisk}) becomes
\begin{equation} \label{eq:disk_radius}
  \rd(t) = 28.0~\mathrm{au}~ \left[ \frac{\eta_\mathrm{A}}{10^{19} \mathrm{cm}^2~\mathrm{s}^{-1}}\right]^{\frac{2}{9}} \left[ \frac{M_\mathrm{acc}(t)}{0.1 \mathrm{M}_\odot}\right]^{\frac{1}{3}} \lambda^{\frac{4}{9}}.
\end{equation}
The ambipolar resistivity $\eta_\mathrm{A}$ depends on the physical and chemical environments. The ambipolar resistivity is of the order of $10^{20}$ cm$^2$ s$^{-1}$ in the envelope, but decreases by a few orders of magnitude in higher density regions such as the disk \citep{Marchand2016}. 
However, the disk radius depends only weakly on the resistivity, with an exponent of 2/9. While it is possible to calculate it self-consistently, it would increase the number of parameters that $\rd$ depends on for a minor gain. Instead, we imposed a value of $\eta_\mathrm{A}=10^{18}$ cm$^{2}$ s$^{-1}$ \citep[consistent with MHD simulations, e.g.,][]{Wurster2016,Wurster2017,Marchand2023,Lebreuilly2023}, so that the user can act more directly on the disk size through the mass-to-flux ratio $\lambda$.

\subsubsection{The disk profile}

We assumed an $\alpha$ disk framework \citep{Shakura1973}, where $\alpha$ is a free-parameter chosen by the user. At a given time, $t$, and cylindrical radius, $\rc$, the viscosity of the gas is given by
\begin{equation}
  \nu(\rc,t) = \alpha \cs(\rc,t) H(\rc,t),
\end{equation}
where
\begin{equation}
  \cs(\rc,t)=\sqrt{\frac{\kb \tm(\rc,t)}{\mu \mp}},
\end{equation}
is the sound speed evaluated at the mid-plane,
\begin{equation}
  H(\rc,t) = \frac{\cs}{\Omk(\rc,t)},
\end{equation}
is the scale height of the disk at a radius $\rc$, and 
\begin{equation} \label{eq:kepler}
  \Omk(\rc,t) = \sqrt{\frac{GM_*(t)}{\rc^3}},
\end{equation}
is the Keplerian angular velocity. 

We used a power-law with an exponential cut-off to describe the surface density profile \citep{Lyndenbell1974}
\begin{equation}\label{eq:disk_sigma}
  \Sigma(\rc,t)=\Sigma_0(t) \left(\frac{\rc}{\rd(t)}\right)^{-1} e^{-\rc/\rd(t)},
\end{equation}
where $\Sigma_0 (t)$ is calculated as a function of the disk mass
\begin{align}
  M_\mathrm{disk}(t) &= \int_{\ri}^{\rd} \int_0^{2\pi} \Sigma(\rc,t) \rc\mathrm{d}\rc\mathrm{d}\phi \nonumber \\
  &= 2\pi \Sigma_0(t) \rd(t)^2 (e^{-\ri/\rd(t)} - e^{-1}).
\end{align}
In APE, the user can choose to either leave the exponential cut-off in eq. (\ref{eq:disk_sigma}), or to sharply cut the disk at a radius $\rd$.
The volume density profile at the mid-plane can then be calculated as
\begin{equation}
  \rho_\mathrm{m} (\rc,t) = \frac{\Sigma(\rc,t)}{H(\rc,t)\sqrt{2\pi}},
\end{equation}
and the vertical profile
\begin{equation}
  \rho(\rc,z,t)= \rho_\mathrm{m} (\rc,t) \exp \left( -\frac{z^2}{2H^2(\rc,t)}\right).
\end{equation}
In addition to the Keplerian angular velocity (eq. \ref{eq:kepler}), the gas in an $\alpha$ disk slowly drifts inward due to the viscosity, with a velocity
\begin{equation}
    v_r (\rc,t) = -\frac{3}{2}\alpha \frac{\cs^2 (\rc,t)}{\rc\Omega_\mathrm{K}(\rc,t)}.
\end{equation}

\subsubsection{The disk temperature}

Many variables depend on the temperature of the disk, through the value of the sound speed. To calculate the temperature of the mid-plane $\tm(\rc,t)$ (dropping the ($\rc$,t) notation hereafter), we followed the method of \citet{Hueso2005}. The temperature value is a balance between the stellar and cloud ambient irradiation, the viscous dissipation of kinetic energy, and the cooling
\begin{equation}\label{eq:disk_temperature}
  \sigma_\mathrm{B} \tm^4 = \frac{1}{8} \left( \frac{3}{8} \tau_\mathrm{R} + \frac{1}{2\tau_\mathrm{P}}\right) \Sigma(\rc,t) \nu(\rc,t) \Omk^2(\rc,t) + \sigma_\mathrm{B} T_\ell^4.
\end{equation}
The first term on the right-hand-side is a sum of optically thin and optically thick contributions \citep{Nakamoto1994}, with $\tau_\mathrm{R}$ and $\tau_\mathrm{P}$ the Rosseland and Planck mean optical depths, respectively.
Similarly to the envelope, $T_\ell$ represents the temperature equilibrium between the stellar irradiation $T_{\ell,1}$ and the ambient cloud temperature $T_\mathrm{mc}=10$ K, so that
\begin{equation}
  T_\ell^4 = T_{\ell,1}^4 + T_\mathrm{mc}^4.
\end{equation}
Temperature $T_{\ell,1}$ is a function of the temperature of the central object $T_*$ and radius $R_*$. It is given by \citep{Adams1988,Ruden1991} and reads
\begin{equation} \label{eq:irradiation}
  T_{\ell,1} = T_* \left[ \frac{2}{3\pi}\left(\frac{R_*}{\rc}\right)^3 + \frac{1}{2}\left(\frac{R_*}{\rc}\right)^2\left(\frac{H}{\rc}\right)\left(\frac{d\ln H}{d \ln \rc}-1\right)    \right]^{\frac{1}{4}},
\end{equation}
where $ d\ln H/d \ln r = 9/7$ is assumed to avoid numerical instability \citep{Hueso2005}.

As \citet{Hueso2005}, we assumed
\begin{equation}
  \tau_\mathrm{P} =2.4 \tau_\mathrm{R},
\end{equation}
and we took
\begin{equation}
  \tau_\mathrm{R} = \frac{\kappa_\mathrm{R} \Sigma}{2}.
\end{equation}
Rosseland opacities $\kappa_\mathrm{R}$ are calculated from the DSHARP opacities \citep{Birnstiel2018} using the local dust grain size-distribution (see Sect. \ref{sec:dust}). The coefficients are averaged over the size-distribution
\begin{equation}
    \kappa_\mathrm{R} = \frac{\int_{a_\mathrm{min}}^{a_\mathrm{max}}n(a)m(a)\kappa_\mathrm{R}(a) \mathrm{d}a}{\int_{a_\mathrm{min}}^{a_\mathrm{max}}n(a)m(a) \mathrm{d}a},
\end{equation}
where $a_\mathrm{min}$ and $a_\mathrm{max}$ are the minimum and maximum grain sizes of the distribution, $n(a)$ the number density of grains of size $a$ and $m(a)$ their mass.
Here, the choice of opacities effectively influences the temperature of the midplane of the disk where the viscosity is high (typically $r\lesssim 30$~au). In turn, this impacts the local scale-height, $H$, and thus the density in the same regions (with subsequent consequences on chemical abundances). For example, higher opacities would increase the temperature and the scale-height, but decrease the density of the gas. The DSHARP opacities are the default in APE, but the user can also generate their own opacity table (the process is explained in the APE user manual provided with the code).

Grains start to sublimate at temperatures $\gtrsim 750$ K, effectively reducing the opacity. Based on \citet{Lenzuni1995} and similarly to \citet{Marchand2016}, we considered that the grain number decreases in several stages, representing the successive sublimation of carbon, silicate and aluminum material. We assumed that grains are made of a unique material, and we modeled this phenomenon by considering a linear decrease in a given material between two temperatures $T_1$ and $T_2$. Those temperatures and relative abundance of the materials are summarized in Table \ref{tab:evaporation}. The temperature at which all grains have sublimated, $T_\mathrm{evap}=1700$ K, is imposed as the maximum temperature in the disk. This impacts only the innermost regions of the disk (typically $\lesssim 5$ au).

\begin{table}
  \caption{Grain sublimation parameters}
  \label{tab:evaporation}
\centering
\begin{tabular}{llll}
\hline\hline
  Material & $T_1$ (K) & $T_2$ (K) & Fraction \\
\hline
 Carbon & 750 & 1100 & 0.883 \\
 Silicate & 1200 & 1300 & 0.112 \\
 Aluminium & 1600 & 1700 & 0.005 \\
\hline
\end{tabular} 
\end{table}

The temperature equation (\ref{eq:disk_temperature}) can then be rewritten in the form
\begin{equation}\label{eq:temperaturetosolve}
  \sigma_\mathrm{B} \tm^4 - \left(Y_1(\rc,t)\kappa_\mathrm{R}(\tm)+\frac{Y_2(\rc,t)}{\kappa_\mathrm{R}(\tm)} \right) \tm - Y_3(\rc,t)\tm^{\frac{1}{2}} - Y_4(\rc,t) = 0,
\end{equation}
where we have
\begin{align}
  Y_1(\rc,t) &= \frac{3}{64} \frac{\kb}{\mu \mp} \alpha \Sigma^2(\rc,t) \Omk(\rc,t) ,\\
  Y_2(\rc,t) &= \frac{5}{96} \frac{\kb}{\mu \mp} \alpha \Omk(\rc,t),\\
  Y_3(\rc,t) &= \sigma_\mathrm{B} T_*^4  \sqrt{\frac{\kb}{\mu \mp}} \frac{1}{7\Omk(\rc,t) \rc} \left( \frac{R_*}{\rc}\right)^2,\\
  Y_4(\rc,t) &= \frac{2}{3\pi} \sigma_\mathrm{B} T_*^4 \left( \frac{R_*}{\rc}\right)^3 + \sigma_\mathrm{B} T_\mathrm{mc}^4.
\end{align}
Solving eq. (\ref{eq:temperaturetosolve}) for $\tm$ provides the mid-plane temperature profile, from which $\cs (\rc,t)$, $H(\rc,t)$, and $\rho(\rc,z,t)$ can be deduced.

\subsection{Outflow}

The magneto-centrifugal forces can launch a collimated jet at the scale of the central object. This creates a conic cavity with an opening angle, $\gamma$, in which low-density gas escapes at velocities of several 10s of km s$^{-1}$.
We assumed that the opening angle varies with height, in accordance to observations \citep{Velusamy1998,Canto2008}. As in \citet{Visser2011}, the cavity boundary is defined by
\begin{equation}
    \rc~\mathrm{(au)} <r_\mathrm{cav}=\left(\frac{z}{0.191~\mathrm{au}}\right)^{\frac{2}{3}} \left(\frac{t}{t_\mathrm{acc}}\right)^2,
\end{equation}
for $t>t_*$, where $t_\mathrm{acc}=2\tff$.
Although the gas velocity in the jet does not affect the global structure, we assumed that this velocity is the escape velocity from the central object taken at the inner radius of the disk
\begin{equation}
    v_\mathrm{esc} = \sqrt{\frac{2GM_*}{r_\mathrm{in}}},
\end{equation}
where 
\begin{equation}
    r_\mathrm{in} = \sqrt{\frac{L_*}{4\pi \sigma_\mathrm{B} T_\mathrm{evap}^4}}.
\end{equation}
The density of the gas inside outflow cavity walls is much lower than that of the disk or the envelope. It is generally measured to be $\sim 10^{3}$ to $\sim 10^6$ cm$^{-3}$ in both models and observations \citep[e.g.,]{Lefloch2015,Rivera-ortiz2023}, with densities of $\sim 10^4$ cm$^{-3}$ generally used in analytical models \citep[e.g.,][]{Visser2009}. Here, we assumed that the outflow density is $10^4$ cm$^{-3}$ at $z=1000$ au, and decreases with height as the cross-section of the cavity increases.
\begin{equation}
    \nh (z) = n_{\mathrm{H},0} \left( \frac{z}{z_0} \right)^{-2},
\end{equation}
with $n_{\mathrm{H},0} = 10^4$ cm$^{-3}$ and $z_0=1000$ au.
The outflow is a low-opacity region through which the radiative flux of the central object can heat the envelope outside the cavity walls. The opening angle therefore impacts the location of the heated region, while the density affects the efficiency of this heating.

\subsection{Dust grains} \label{sec:dust}

We included a treatment of the dust size-distribution to estimate the opacities for radiative transfer calculations. While the characteristic size of grains is sub-micron in the interstellar medium \citep{Mathis1977}, they grow by coagulation during star formation and can quickly reach radii larger than 100 $\mu$m in protoplanetary disks \citep{Marchand2023}. In APE, we have implemented the simulation results of \citet{Lebreuilly2023}. They performed 1D non-ideal MHD simulations of protostellar collapses using the SHARK code \citep{Lebreuilly2023} to track the evolution of the grain size-distribution by coagulation and fragmentation. This evolution is controlled by the collision rates between grains. \citet{Lebreuilly2023} take into account differential grain-grain velocities originating from the gas turbulence \citep{Ormel2007}, ambipolar diffusion \citep[due to the electrical charges of grains, see][]{Guillet2020} and Brownian motion \citep{Bate2022}. Turbulence is efficient at growing large grains in high-density environments (as the disk), while ambipolar diffusion and Brownian motion efficiently remove small grains in the envelope. In the disk, the fragmentation limits the growth and repopulate the small-grains population.

\begin{figure}
    \centering
    \includegraphics[width=0.5\textwidth, trim=0cm 0cm 0cm 0cm]{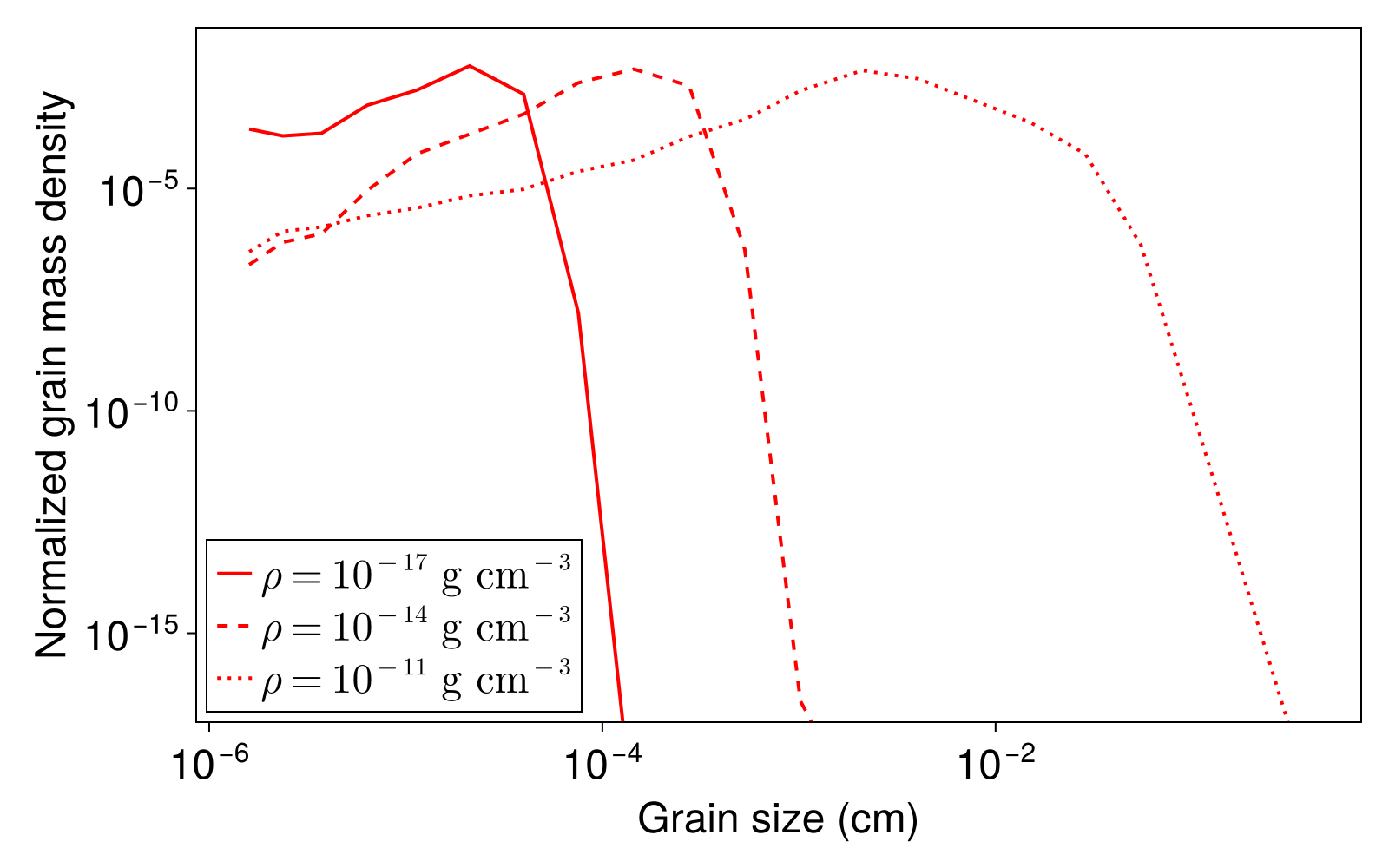}
    \caption{Grain size-distributions at various densities. The y axis represents the mass density of the grains normalized by the mass density of the gas.}
    \label{fig:ape_size-dist}
\end{figure}

We used the last output of the simulation PC1-FRAG of \citet{Lebreuilly2023} to associate gas densities with size-distributions. The simulation was originally performed using 100 bins in size between 7 nm and 1 cm, with a dust-to-gas mass ratio of 1\%. This large number of bins may however result in expensive radiative transfer calculations, and we resampled the distribution into 20 bins between 16 nm and 3.9 mm. In APE, we provide the resampled distribution and the associated opacities for radiative transfer post-processing. Size-distributions at $\rho=10^{-17}$, $10^{-14}$ and $10^{-11}$ g~cm$^{-3}$ are displayed in Fig. \ref{fig:ape_size-dist}. In the envelope (represented by the solid line), the distribution is close to the initial Mathis, Rumpl, Nordsieck (MRN) distribution \citep{Mathis1977}, with the peak of the distribution shifting toward larger sizes as the density increases.

The results of \citet{Lebreuilly2023} are however only valid for a dust-to-gas mass ratio of 0.01. As users may desire different dust-to-gas mass ratios, we provide an alternative grain coagulation model. \citet{Marchand2021} show that, when considering only collisions from turbulence, the dust coagulation can be treated as a 1-dimension problem. Size-distributions depend then only on the initial distribution, and one variable that they name $\chi$, which is a function of the density and temperature history of the grains. This variable also scales linearly with the dust-to-gas mass ratio. 3D simulations with non-ideal MHD by \citet{Marchand2023} present the evolution of $\chi$ and the dust size throughout the collapse. From their simulation, we derived a prescription for $\chi$ at a dust-to-gas mass ratio of 0.01
\begin{equation}
     \chi (0.01) = \left\{ \begin{tabular}{lr}
        $10^{16} \left( \frac{\rho}{3\times 10^{-20}} \right)^{\frac{1}{4}}$ &  $(\rho\leq 10^{-15}~\mathrm{g~cm}^{-3})$\\
        $1.34 \times 10^{18} \left( \frac{\rho}{3.83 \times 10^{-12}} \right)^{\frac{1}{15}} \left( \frac{T}{10} \right)^{\frac{1}{15}}$&  $(\rho > 10^{-15}~\mathrm{g~cm}^{-3})$
    \end{tabular}
    \right. .
\end{equation}
$\chi$ is then linearly scaled by the desired dust-to-gas mass ratio, $dtg$.
\begin{equation}
    \chi (dtg) = \chi (0.01) \frac{dtg}{0.01}.
\end{equation}
The prescription is discontinuous as the growth of grains is fast at densities $\rho >10^{-15}$ g cm$^{-3}$ \citep{Marchand2023}. The outflow is composed of material lifted from regions in the vicinity of the disk, typically at densities close to $\rho = 10^{-15}$ g cm$^{-3}$ \citep{Marchand2023}; we therefore considered that the value of $\chi$ in the outflow matches the one at density $\rho = 10^{-15}$ g cm$^{-3}$.
We have generated a tabulated size-distribution evolution path as a function of $\chi$ using the Ishinisan code \citep{Marchand2021}. We provide this evolution path with APE. The dust abundance is then read from the table using the local values of $\chi$.

In APE, choosing a dust-to-gas mass ratio exactly equal to 0.01 is equivalent to choosing the coagulation model of \citet{Lebreuilly2023}. Any other value of $dtg$ switches the coagulation model to \citet{Marchand2023}. The user can also choose to turn off the coagulation to only have a MRN distribution. 

As grains grow, their total surface area decreases, which can impact the chemical evolution. However, to our knowledge, no chemical code currently supports an evolving grain size-distribution. Nautilus, in particular, supports only a fixed grain size with a fixed dust-to-gas mass ratio. A few attempts at emulating the evolution of grains in chemical networks have been published. For example, \citet{Iqbal2018} implemented a fixed, full size-distribution in Nautilus. They found that differences in the chemical evolution between two size-distributions could be small as long as the total surface areas were close, but could become significant otherwise or in the presence of cosmic-rays. More recently, \citet{Navarro2024} modified Nautilus by computing an effective grain size at each step. They found that the main impact of the decrease in grain surface area is seen in dense regions, below 100~K, as many species are still in solid phase.
APE has been designed to interface with any chemical code, preferentially with the public version of Nautilus. Tools and methods to account for an evolving grain size-distribution in chemical simulations are still being developed and, to this day, none are publicly available. Nonetheless, this is a crucial aspect of the chemistry of star formation, and APE is able to provide evolving size-distributions to future codes that include this physics. Until such implementations become available, we do not account for grain growth in our chemical runs.

\subsection{Extinction}

APE has been designed to work with the three-phase kinetic chemistry code Nautilus \citep{Ruaud2016} by generating necessary input files. Nautilus accounts for the impact of external UV emission on the chemical species, and requires an extinction value. We estimated the extinction, $A_\mathrm{v}$, through the column density $N_{\mathrm{H}_2}$ of the gas between the point of interest and the outside of the cloud \citep{Semenov2010}
\begin{equation}
    A_\mathrm{v}=\frac{N_{\mathrm{H}_2}}{1.59 \times 10^{21}~\mathrm{g}~\mathrm{cm}^{-2}}.
\end{equation}
We assumed a power-law density profile in the envelope, $\rho(r)\sim r^{s}$ (with $s$ converging toward -1.5, see Sect. \ref{sec:envelope_density}). After estimating $s$, an integration is performed toward the outer-edge of the envelope to obtain the column density
\begin{align}
    N_{\mathrm{H}_2,\mathrm{env}} (r) &= \int_r^{r_\mathrm{max}} \rho(r')\mathrm{d}r',\\
    & = \frac{\rho(r)}{s+1}\frac{r_\mathrm{max}^{s+1}-r^{s+1}}{r^{s}},\label{eq:NH2_env}\\
    &= \exp{}\left[ \ln(\rho(r)) +\ln(r) -\ln(-(s+1)) +\ln\left(1-\left(\frac{r_\mathrm{max}}{r}\right)^{s+1} \right) \right].\label{eq:NH2_env_2}
\end{align}
Equations (\ref{eq:NH2_env}) and (\ref{eq:NH2_env_2}) are equivalent, but we employed the latter in APE to avoid numerical precision issues. We have added a negative sign to the last two terms of eq. \ref{eq:NH2_env_2} because $s+1<0$.

In the disk, for the gas at cylindrical radius $\rc$ and height $z$, the density profile is integrated vertically until one scale-height (dropping the dependency in $t$):
\begin{align}
  N_{\mathrm{H}_2,\mathrm{disk}} (\rc,z)  &= \int_z^{H(\rc)} \rho_\mathrm{m}(\rc) \exp\left( -\frac{z'^2}{2H^2(\rc)}\right) \mathrm{d}z',\\
  & = \sqrt{\frac{\pi}{2}} \rho_\mathrm{m}(\rc) H(\rc) \left[ 1- \mathrm{erf}\left( \frac{z}{\sqrt{2}H(\rc)}\right) \right],
\end{align}
with $\mathrm{erf}$ the error function.
We added a flat value to account for the external cloud extinction. This value is a parameter, which can be chosen by the user. 
Finally, the total extinction is the sum of the three contributions
\begin{equation}
    A_\mathrm{v}(r,\rc,z) = A_\mathrm{v,env}(r) + A_\mathrm{v,disk}(\rc,z) + A_\mathrm{v,cloud}.
\end{equation}

\subsection{APE workflow}

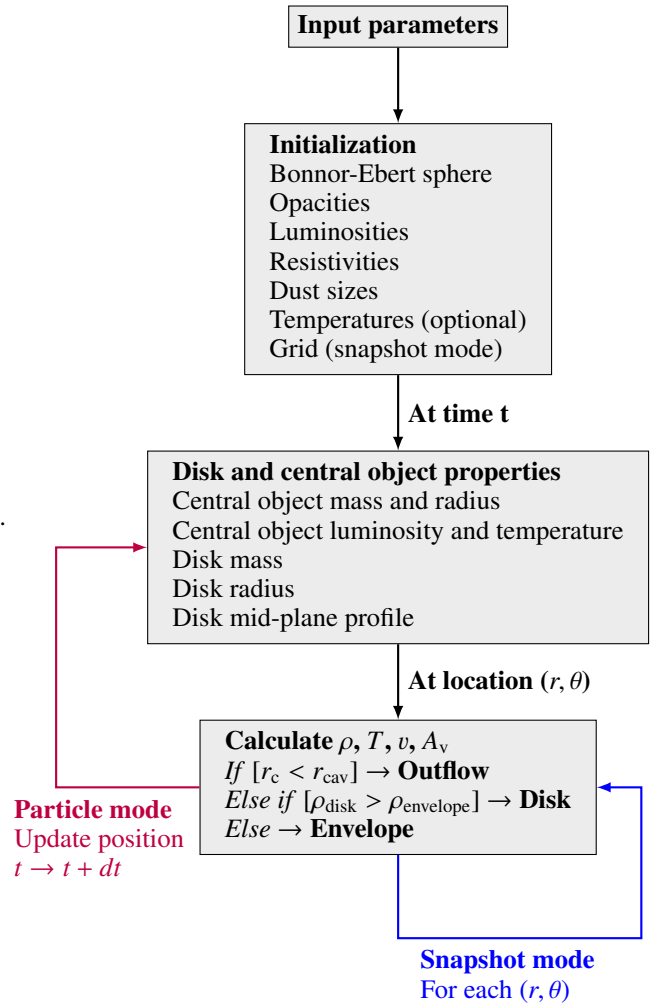
\begin{figure}
    \centering
\begin{tikzpicture}
    \node[nodebox] (Inp) at (0,0) {\textbf{Input parameters}};
    
    \node[nodebox, below = of Inp] (Ini) {\begin{tabular}{l} \textbf{Initialization}\\ Bonnor-Ebert sphere \\ Opacities \\ Luminosities \\ Resistivities \\ Dust sizes \\ Temperatures (optional) \\ Grid (snapshot mode) \end{tabular}};
    
    \node[nodebox, below=of Ini] (Props) {\begin{tabular}{l} \textbf{Disk and central object properties}\\ Central object mass and radius \\ Central object luminosity and temperature \\ Disk mass \\ Disk radius \\ Disk mid-plane profile \end{tabular}};
    
    \node[nodebox, below = of Props] (Calc) {
    \begin{tabular}{l} \textbf{Calculate $\rho$, $T$, $v$, $A_\mathrm{v}$}\\
    \textit{If} $[\rc < r_\mathrm{cav}]$ $\rightarrow$ \textbf{Outflow} \\
    \textit{Else if} $[\rho_\mathrm{disk} > \rho_\mathrm{envelope}]$ $\rightarrow$ \textbf{Disk} \\
    \textit{Else} $\rightarrow$ \textbf{Envelope} \\
    \end{tabular}};
    
    \draw[arrow] (Inp) -- (Ini);
    \draw[arrow] (Ini) -- (Props) node[pos=0.5,right, anchor=west] {\textbf{At time t}};
    \draw[arrow] (Props) -- (Calc) node[pos=0.5,right, anchor=west] {\textbf{At location ($r,\theta$)}};
    
    \draw [purple,arrow] (Calc) -- node[pos=0.7,below, anchor=north] {\begin{tabular}{l} \textbf{Particle mode}\\Update position \\ $t \rightarrow t+dt$ \end{tabular}} node {} ++(-4.5cm,0) |-  (Props)  ;
    
    \draw [blue,arrow] (Calc) --  node {} ++(0cm,-2cm) -- node {} ++(3.2cm,0cm) node[pos=0.45, below, anchor=north] {\begin{tabular}{l} \textbf{Snapshot mode}\\For each ($r,\theta$) \end{tabular}} |-  (Calc)  ;
\end{tikzpicture}
    \caption{Workflow of the APE code.}
    \label{fig:ape_workflow}
\end{figure}

Unlike hydrodynamics simulations, determining the physical properties at a specific location in APE does not require the description of the whole history of the gas. For instance, the density at location ($r$,$\theta$) at time $t$ is calculated with no knowledge of the gas surrounding this location or the system at earlier time points. The calculation solely relies on key characteristics of the central object (mass, radius, luminosity, temperature) and the disk (mass, size, mid-plane profile). This feature significantly accelerates the computation: a snapshot at time $t$ can be generated without computing the whole gas history since $t=0$, and a particle trajectory can be traced without computing the whole map at each time-step. Hence, the grid resolution does not influence the physical result (although it impacts the post-processing calculations like radiative transfer and synthetic observations).
The APE workflow is summarized in Fig. \ref{fig:ape_workflow}.

\subsection{The trajectory of particles} \label{sec:f_factor}

\begin{figure}
    \centering
    \includegraphics[width=0.5\textwidth, trim=0cm 0cm 0cm 0cm]{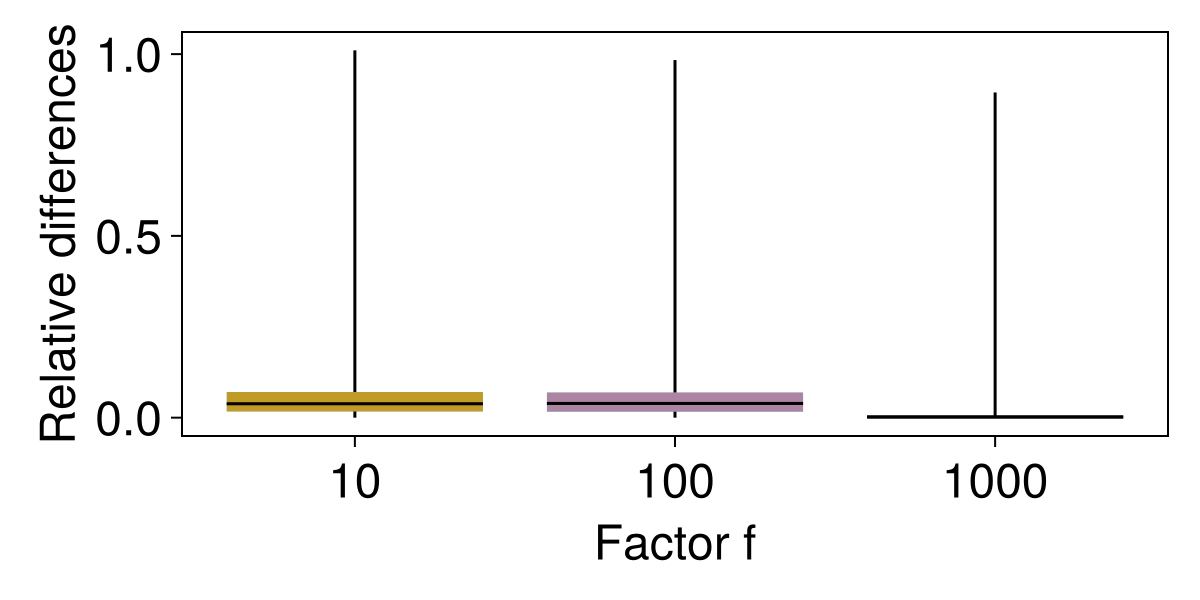}
    \caption{Distributions of the relative differences in abundances for several values of $f$, using $f=10000$ as reference. The horizontal black lines represent the medians of the distributions, the colored boxes range from the first to the third quartiles, and the vertical black lines span from the minimum to the maximum values. For the $f=1000$ case, the colored box is not visible as its upper limit stands at a value of 0.001.}
    \label{fig:convergence_chem}
\end{figure}

In particle mode, particles start with an initial user-defined position and move according to the local velocity field. The default time-step is $dt_0=200$ years (this value can be changed by the user). However, the time-step is limited when the velocity of the particle is high or if the particle is close to the edges of the simulation box. The time-step is calculated as
\begin{equation}
    dt = \mathrm{min}\left(\frac{1}{f}\frac{r}{v_r}, dt_0\right), 
\end{equation}
with $r$ the spherical radial coordinate, $v_r$ the radial velocity, and $f$ an arbitrary safety factor. Dynamically, this factor is especially important in the inner envelope, where radial velocities can reach several km s$^{-1}$. Conversely, the radial velocity in the disk is low ($\sim 10^{-3}$ km s$^{-1}$), and the particle trajectory is little sensitive on the factor $f$.
However, the value of the time-step may influence the chemical evolution of the particle. We performed a convergence study on the influence of $f$ upon chemical abundances. We ran APE with a particle starting at position ($x=\rc=5$ au, $z=1$ au) (in the disk) at $t=\tff+40$ kyr, in reverse mode until $t=0$, with $dt_0=200$ years. The system parameters are the same as the fiducial case presented in Sect. \ref{sec:simulation_setup}, with a cloud mass of $2$ M$_\odot$ and a mass-to-flux ratio $\lambda=5$. The particle spends $\sim 150$ kyr in the envelope and $\sim 40$ kyr in the disk. This final position has be chosen so that the particle travels across a wide range of physical conditions, from the envelope to the inner disk, and is thus more impacted by changes in the time-step. We investigated values of $f$ of 10, 100, 1000 and 10000. Dynamically, the trajectories are very close, with less than a 0.4\% difference in their final positions, as listed in Table \ref{tab:chem_convergence}. We ran the Nautilus code on those trajectories using the publicly provided chemical network. In Fig. \ref{fig:convergence_chem}, we display box plots of the relative differences of abundance for all chemical species in the network, using $f=10000$ as the reference case. For $f$=10 and 100, half of the species have less than a 4\% difference with respect to the reference case. Over 95\% of species have less than a 15\% difference. Nearly all species have less than a 50\% difference, except 4 species in gas and solid phase for $f$=10 (FeOH, HFeOH, HOFeOH, and FeHOHOH), and 2 species for $f$=100 (NH$_2$PO and HOFeOH), while all within a factor 2. The case $f$=1000 is well converged with all species displaying less than 5\% difference with $f$=10000, except HCONO with a 8\% difference and NH$_2$PO with a 90\% difference. For chemical convergence, we therefore recommend to use $1000 \geq f > 100$, but choosing $f=10$ results in a $\sim$50\% maximum error on abundances while requiring much less computational resource. We reach the same conclusions for a particle starting at (x=1 au, z=5 au) and staying exclusively in the envelope.

\begin{table}
  \caption{APE computation results for various factors $f$ limiting the time-step of the particles.}
  \label{tab:chem_convergence}
\centering
\begin{tabular}{llll}
\hline\hline
  $f$ & Number of time-steps & Initial position & Time $^a$\\
\hline
 10 & 981 & (6100 au, 991 au) & 6.6 s\\
 100 & 1317 & (6090 au, 989 au) & 7.5 s\\
 1000 & 7108 & (6083 au, 988 au) & 49 s\\
 10000 & 70975 & (6081 au, 987 au) & 490 s\\
\hline
\end{tabular}
\tablefoot{$^a$ APE particle computation times, performed in 2024 on a 5 year old laptop. The times for a particle ending close to the center at ($x=\rc=5$ au, $z=1$ au) and staying in the envelope are up to 5 times larger, due to more stringent constraints on the time-step.}
\end{table}

\subsection{Producing synthetic observations}\label{sec:production}

The purpose of APE is to generate physical conditions to run chemical models and synthetic observations. Producing synthetic observations requires simulating the telescope view of an emission map, which is itself computed by radiative transfer calculations from abundance maps. While the users of APE are free to use any software to perform those steps, we only provide scripts to the publicly available codes Nautilus \citep{Ruaud2016}, RADMC-3D \citep{Dullemond2012}, and Imager\footnote{\url{http://www.iram.fr/IRAMFR/GILDAS}\\ \url{https://imager.oasu.u-bordeaux.fr}} to facilitate the process.

For molecular line observations, the first step is the production of chemical abundance maps. The ``grid of particles'' mode of APE places virtual particles in each cell of the map at a given time $t_\mathrm{final}$ (generated in the same way as in a snapshot), and computes their trajectory backward in time until the initial condition $t=0$ is reached. The history of each particle can then be used as input for Nautilus, alongside initial abundances and a chemical reaction network, to compute its chemical evolution. The chemical abundance maps can then be reconstructed into a grid, similarly to a density snapshot, from the final abundances of all particles at time $t_\mathrm{final}$. 

The next step is the computation of emission maps. To that end, RADMC-3D needs information about the grid data (automatically generated by APE), spectroscopic data (user-provided) and observational parameters. Those parameters are user-defined and include the observation wavelength (alternatively which molecular transitions to observe), and the source inclination. RADMC-3D then reconstructs the 3D spatial distribution of the molecule from the axisymmetric APE data, and produces an emission cube. 
Finally, Imager is run on the emission map with the ``simulate'' function to produce cubes of observations.

With APE, we provide a script, available both in Python and Julia, in which the user can choose all observation parameters: the species, the frequency bandwidth, the spectral resolution, the inclination, the source distance and declination, the ALMA configuration, the observing time, and the noise. The script is an interface between the outputs of APE (once the chemical abundance maps have been generated), RADMC-3D, and Imager. The final output is a FITS file with the clean cube, optionally in which the continuum has been subtracted.

\section{Application}\label{sec:application}

\subsection{Computing abundance maps}\label{sec:application_chem}

\subsubsection{Method}

In this section, we illustrate an application of APE, which is to compute the abundance map of several molecules, using the ``Grid of particles'' mode described in Sect. \ref{sec:production}. The temperature is interpolated in time and space along the trajectory of particles using 
RADMC-3D as explained in Sect. \ref{sec:radiative_transfer}.

\subsubsection{Simulation setup}\label{sec:simulation_setup}

We considered a cloud with an initial mass of 2 M$_\odot$, and produce a snapshot 150 kyr after the formation of the central object, which occurs at $t=\tff \approx 152$ kyr.
The disk size is determined by the mass-to-flux ratio, which we chose to be $\lambda=5$ as it is a standard value in protostellar collapse simulations \citep[e.g.,][]{Masson2016,Wurster2016,Marchand2023}. The dust-to-gas mass ratio is 1\% and we enabled dust coagulation. The initial angular velocity of the cloud is set at $\Omega_0=2\times 10^{-15}$~s$^{-1}$.
The grid is composed of 5625 cells, with a logarithmic radial sampling of 75 points between $r=1$ au and $r=1000$ au, and a uniform poloidal sampling of 75 points between $\theta=0$ and $\theta=\pi/2$. We used a limiting factor $f=300$ on the time-step (see Sect. \ref{sec:f_factor}).

\subsubsection{Chemical model}\label{sec:chem_model}

Each particle history is used as input for the Nautilus code \citep{Ruaud2016}, which performs three-phase gas-grain chemical calculations (gas phase - grain surface - grain mantle). 
The chemical network used in the Nautilus code contains about 800 species and $\sim 9000$ individual reactions \citep{RN10861, RN11273, RN11591}. The grain surface and the mantle are both chemically active for these simulations. A sticking probability of 1 is assumed for all neutral species while desorption can occur by thermal and non-thermal (cosmic-ray, chemical desorption) processes including sputtering of ices by cosmic-ray collisions \citep{RN11329}. Surface reactions formalism and more detailed description of the simulations can be found in \citet{RN6446}. The diffusion energies are set to be a fraction of the binding energy for the surface (0.4 times) and the mantle (0.8 times), with a diffusion barrier thickness of 0.25 nm.

Initially, we ran Nautilus for $t_\mathrm{pre}=10^6$ years under constant physical parameters characteristic of prestellar core conditions: a density of $n_0=10^4$ cm$^{-3}$, a temperature of $T_0=10$ K, and an extinction of $A_\mathrm{v}=5$. We assumed a cosmic-ray ionization rate of $\zeta_\mathrm{CR}=1.3\times 10^{-17}$ s$^{-1}$. We started this run using the same abundances as \citet{Ruaud2018}, which we detail in Table \ref{tab:ini_abundances} (with the exception of Mg that is not included in the chemical network).
In the rest of the paper, we call “initial abundances” the final abundances from this run, which serve as the initial conditions for the chemical simulations along each particle trajectory.
Since Nautilus does not support yet an evolving grain size-distribution \citep[although a first attempt has been recently published by][]{Navarro2024}, we used a unique, constant grain size that we assumed to be 0.1 $\mu$m, with a dust-to-gas mass ratio of 0.01. The bulk density of the grains is 3 g cm$^{-3}$, and their surface site density of $1.5 \times 10^{15}$ cm$^{-2}$.
In this work, we did not compute the chemical abundances in the outflow cavity, where they are set equal to the initial abundances. In this study, we considerd that the sole impact of the outflow is on temperature maps, which can in turn affect chemical abundances.

\begin{table}
  \caption{Initial atomic abundances.}
  \label{tab:ini_abundances}
\centering
\begin{tabular}{llll}
\hline\hline
  Species & Abundance (/$n_{\mathrm{H}}$) & Species & Abundance (/$n_{\mathrm{H}}$)\\
\hline
 H & 1 & Si$^+$ & $1.8 \times 10^{-6}$ \\
 He & 0.09 & Fe$^+$ & $2.0 \times 10^{-7}$\\
 N & $6.2 \times 10^{-5}$ & Na$^+$ & $2.3 \times 10^{-7}$\\
 O & $3.3 \times 10^{-4}$ & P$^+$ & $7.8 \times 10^{-8}$\\
 C$^+$ & $1.8 \times 10^{-4}$ & Cl$^+$ & $3.4 \times 10^{-8}$\\
 S$^+$ & $1.5 \times 10^{-5}$ & F & $1.8 \times 10^{-8}$\\
\hline
\end{tabular} 
\end{table}

\subsubsection{System evolution}

\begin{figure}
    \centering
    \includegraphics[width=0.5\textwidth, trim=5cm 0cm 0cm 0cm,clip]{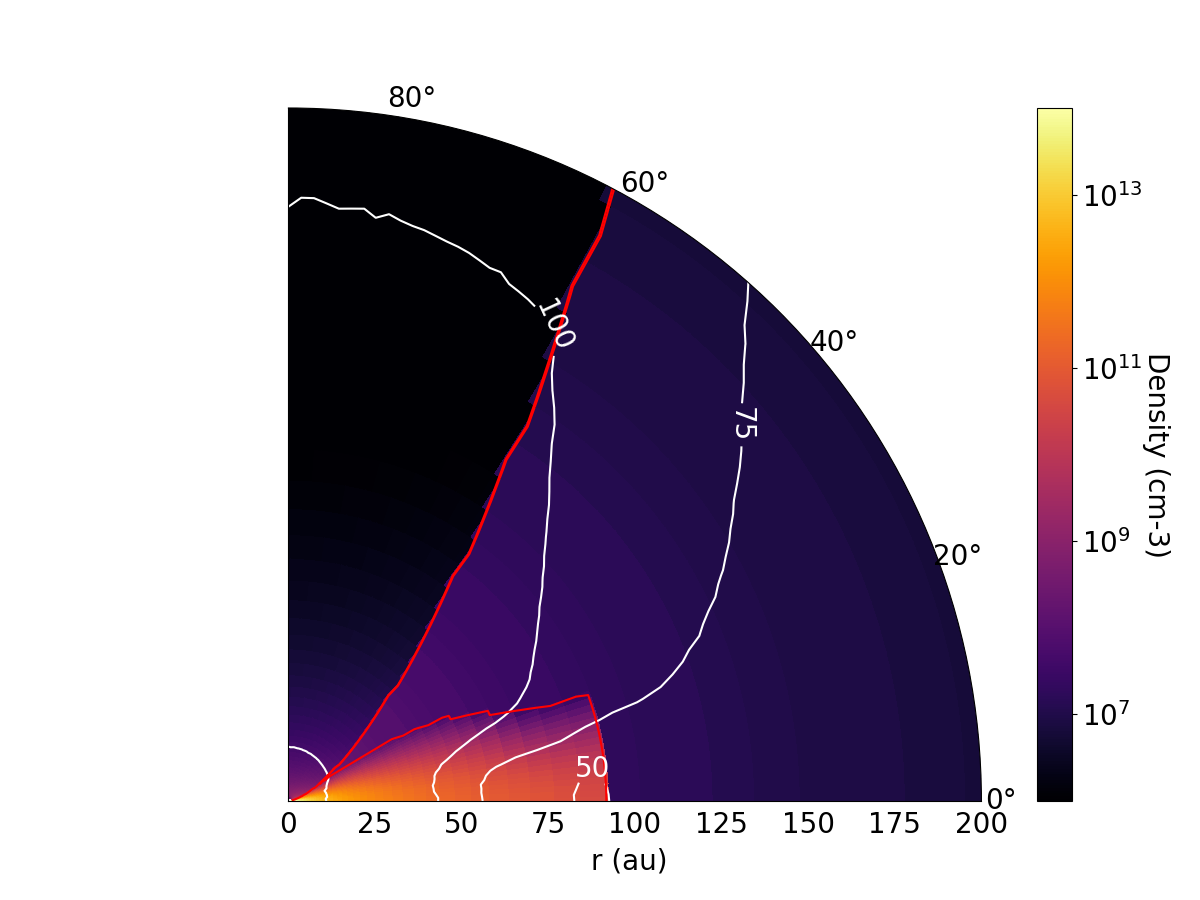}
    \caption{Zoomed map of the system at $t=\tff+$150 kyr. Colors represent the number density of the gas, while white lines indicate iso-contours of the temperature at T= 50, 75, 100 and 300 K. The red contours delimitate the envelope, the disk and the outflow.}
    \label{fig:density_temp_time}
\end{figure}

At $t=\tff+150$ kyr, the envelope has nearly vanished with a remaining mass of $0.05$ M$_\odot$. The central object has accreted $1.5$ M$_\odot$, while the disk has the remaining mass of 0.45 M$_\odot$, and a radius of 92.3 au. Figure \ref{fig:density_temp_time} displays the density and temperature map of the system zoomed on the inner 200 au. The protostellar radiation hardly penetrates into the outer regions of the disk, resulting in a local minimum of temperature.

\begin{figure}
    \centering
    \includegraphics[width=0.5\textwidth, trim=0cm 0cm 0cm 0cm]{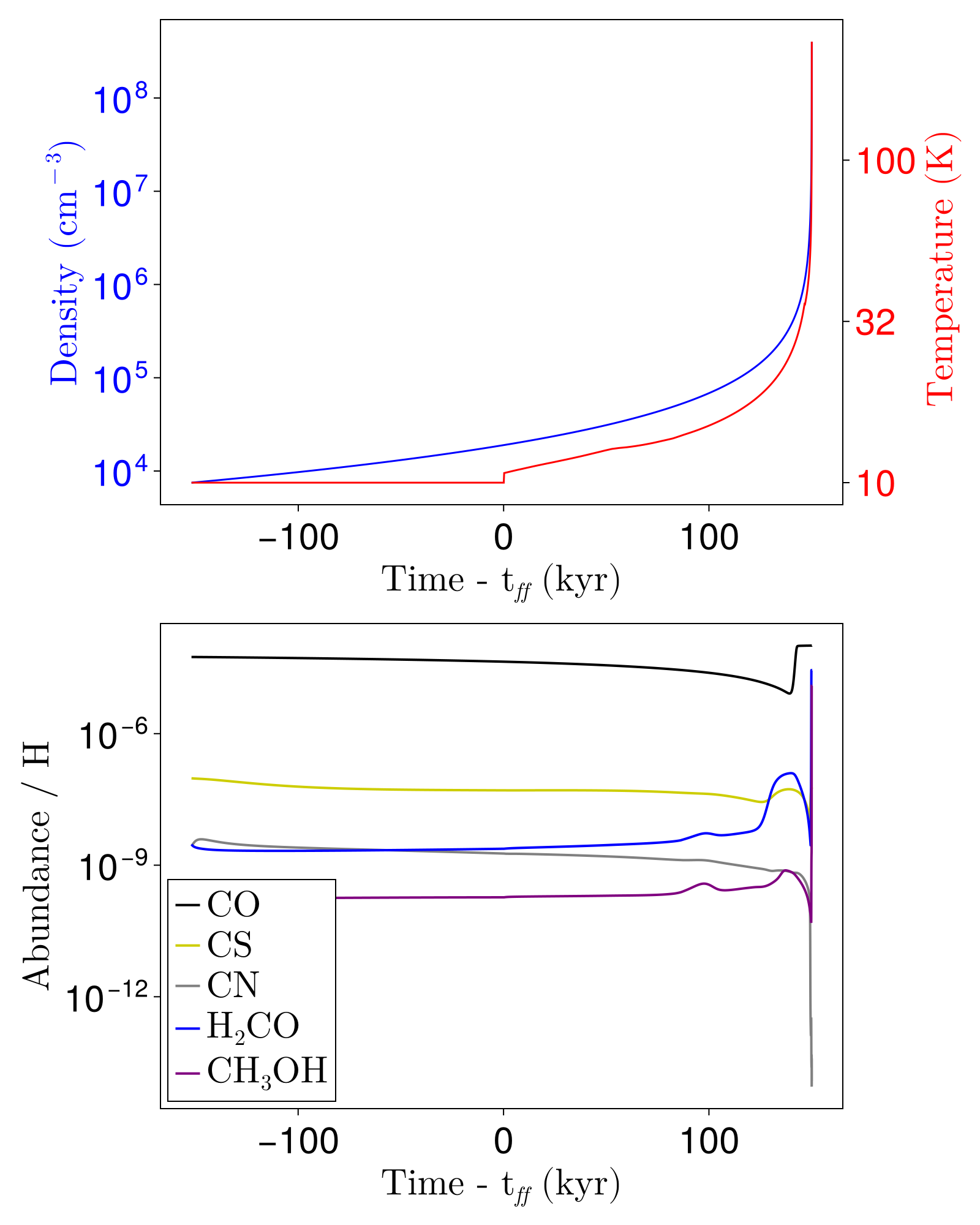}
    \caption{Upper panel: Density (blue) and temperature (red) history as a function of time for a particle ending in the inner envelope ($x,y = 14.0,14.0$ au) at its last time-step at $t=\tff + 150$~kyr. Bottom panel: Time evolution of the gas phase abundance of several species for this particle.}
    \label{fig:part_traj_150_env}
\end{figure}

\begin{figure}
    \centering
    \includegraphics[width=0.5\textwidth, trim=0cm 0cm 0cm 0cm]{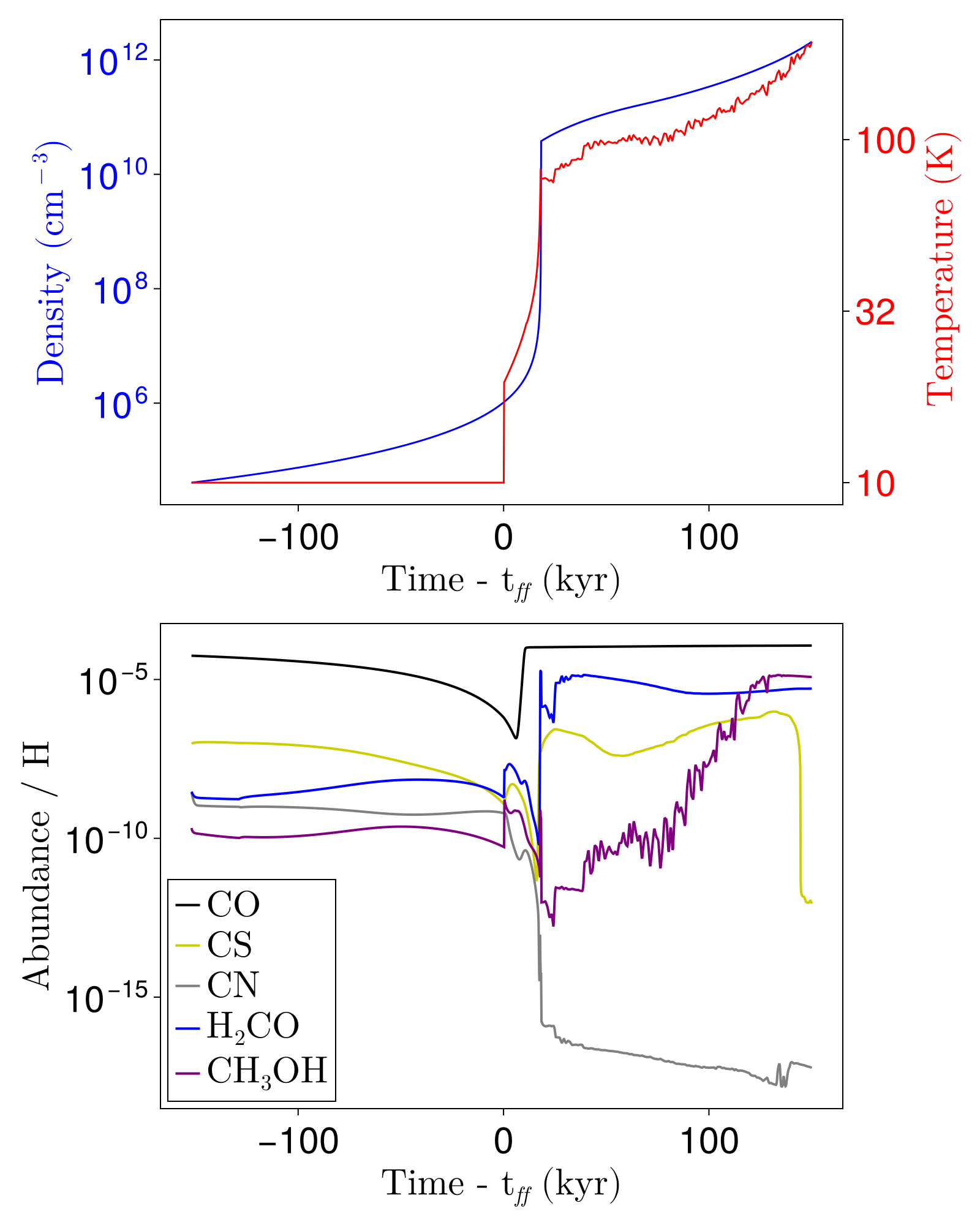}
    \caption{Same as Fig. \ref{fig:part_traj_150_env} for a particle ending in the disk ($x,y = 19.1,2.0$ au).}
    \label{fig:part_traj_150_disk}
\end{figure}

The upper panel of Fig. \ref{fig:part_traj_150_env} displays the density and temperature history of a particle located in the envelope at a distance of $\sim 20$ au from the central object at the final time of $t=\tff+150$~kyr. Figure \ref{fig:part_traj_150_disk} is the same for a particle ending in the disk.
The envelope particle starts at a density of $\sim 10^{4}$~cm$^{-3}$, which increases only by one order of magnitude over 250~kyr. The density increase then proceeds faster as the particle draws close to the central object, to reach $\sim 10^{8}$~cm$^{-3}$ at the end of the simulation.
The density of the disk particle starts at $\sim 10^5$ cm$^{-3}$ and slowly increases to $\sim 10^{7}$ cm$^{-3}$, until it reaches the inner envelope ($\lesssim 100$ au) at the time of formation of the central object ($t-\tff=0$). This particle then enters the disk and experiences a density jump, before slowly drifting inward until the final time.
The temperature of both particles is initially constant at $T=T_\mathrm{mc}=10$ K. The particles are then close enough to the central object at its formation to experience a jump of temperature, at $t-\tff=0$~kyr. The disk particle immediately reaches $T \approx 20$ K, then heats up to $T\approx 80$ K at its entry in the disk. Its temperature then increases slowly while drifting inward, reaching $T\approx 120$~K at the end of the simulation. The envelope particle is further away at the formation of the central object, and its temperature only jumps to $T\approx 12$ K, and then heats up increasingly faster the following 150 kyr until it also reaches $T \approx 120$ K.

\subsubsection{Chemical abundances} \label{sec:abundance}

The density and temperature evolution are then used as input for the chemical calculations. As illustration, the bottom panels of Figs. \ref{fig:part_traj_150_env} and \ref{fig:part_traj_150_disk} show the time evolution of the abundance (relative to H) of several species: carbon monoxyde (CO), carbon monosulfide (CS), cyanide (CN), formaldehyde (H$_2$CO), and methanol (CH$_3$OH) in the gas phase for the same particles as in the upper panels. The chemical evolution of the envelope particle (Fig. \ref{fig:part_traj_150_env}) is slow, with variations smaller than one order of magnitude, until a few 10~kyr before the end of the simulation. Close to the protostar, the abundances then vary rapidly: the abundance of CN decreases by 5 orders of magnitude, while the abundance of other molecules increase by 1 to 4 orders of magnitude.
The disk particle (Fig. \ref{fig:part_traj_150_disk}) starts its journey closer to the center of the cloud. At the formation of the central object, the jump in temperature causes a $\sim 1$ order of magnitude increase in the gas phase abundance of H$_2$CO and CH$_3$OH, a decrease by the same factor for CN, while CO and CS are little affected. CO sees a decrease in abundance until the disk entry, where it suddenly increases by 2 orders of magnitude and stays constant until the end. H$_2$CO and CS exhibit a similar behavior with jumps of 4 orders of magnitude. In the last kyrs of the simulation, the abundance of CS sharply decreases by 5 orders of magnitude. Conversely, the abundance of CN drops by 8 orders of magnitudes at the disk entry, and continues decreasing by 2 orders of magnitude over the following $\sim 100$ kyr. The density jump at the disk entry also causes a sudden decrease in CH$_3$OH gas phase abundance by 2 orders of magnitude. As the particle drifts inward in the disk, the increasing temperature provokes a slow desorption of CH$_3$OH in the gas phase over time. At the final time, all CH$_3$OH molecules present on grains have desorbed and their abundance is 5 orders of magnitude higher than before the disk entry.

\begin{figure*}
    \centering
    \includegraphics[width=0.32\textwidth, trim=6cm 0cm 0cm 0cm,clip]{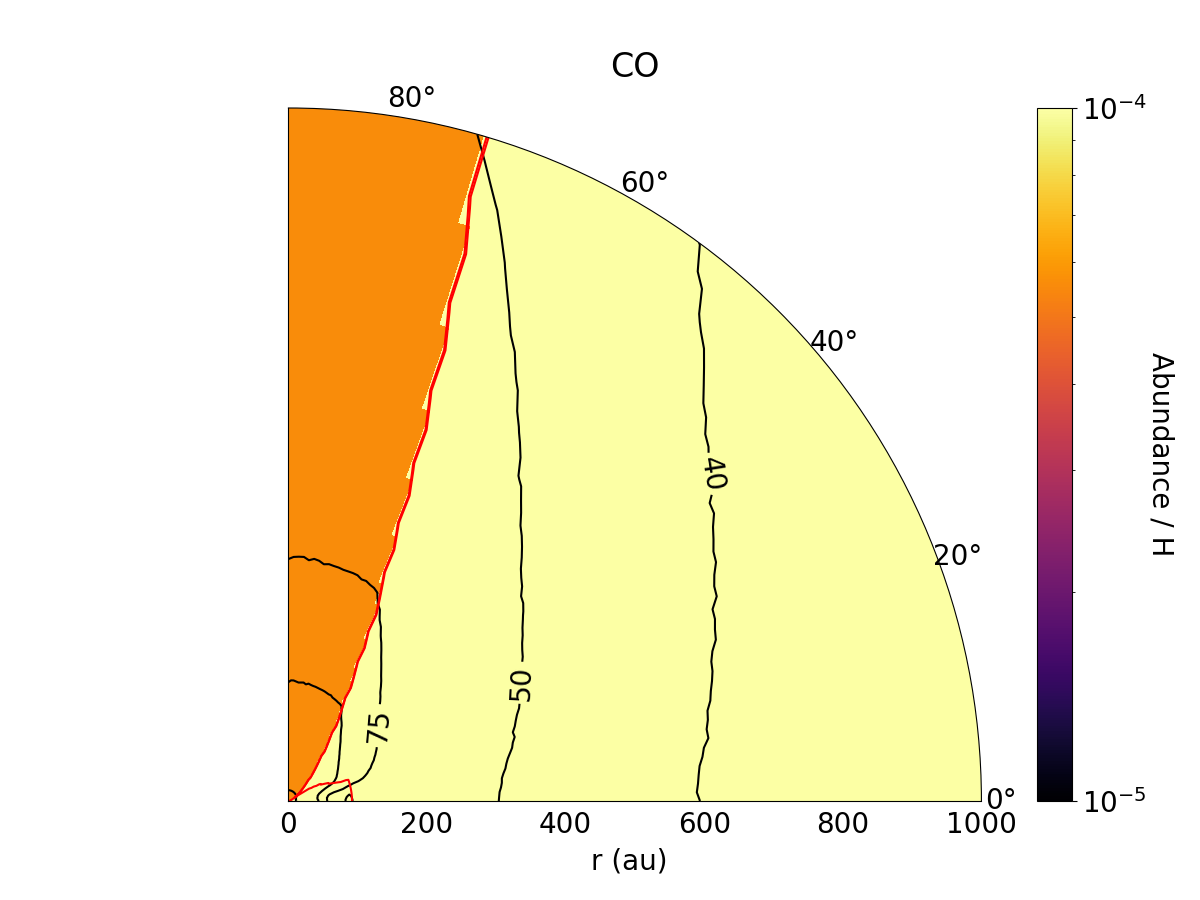}
    \includegraphics[width=0.32\textwidth, trim=6cm 0cm 0cm 0cm,clip]{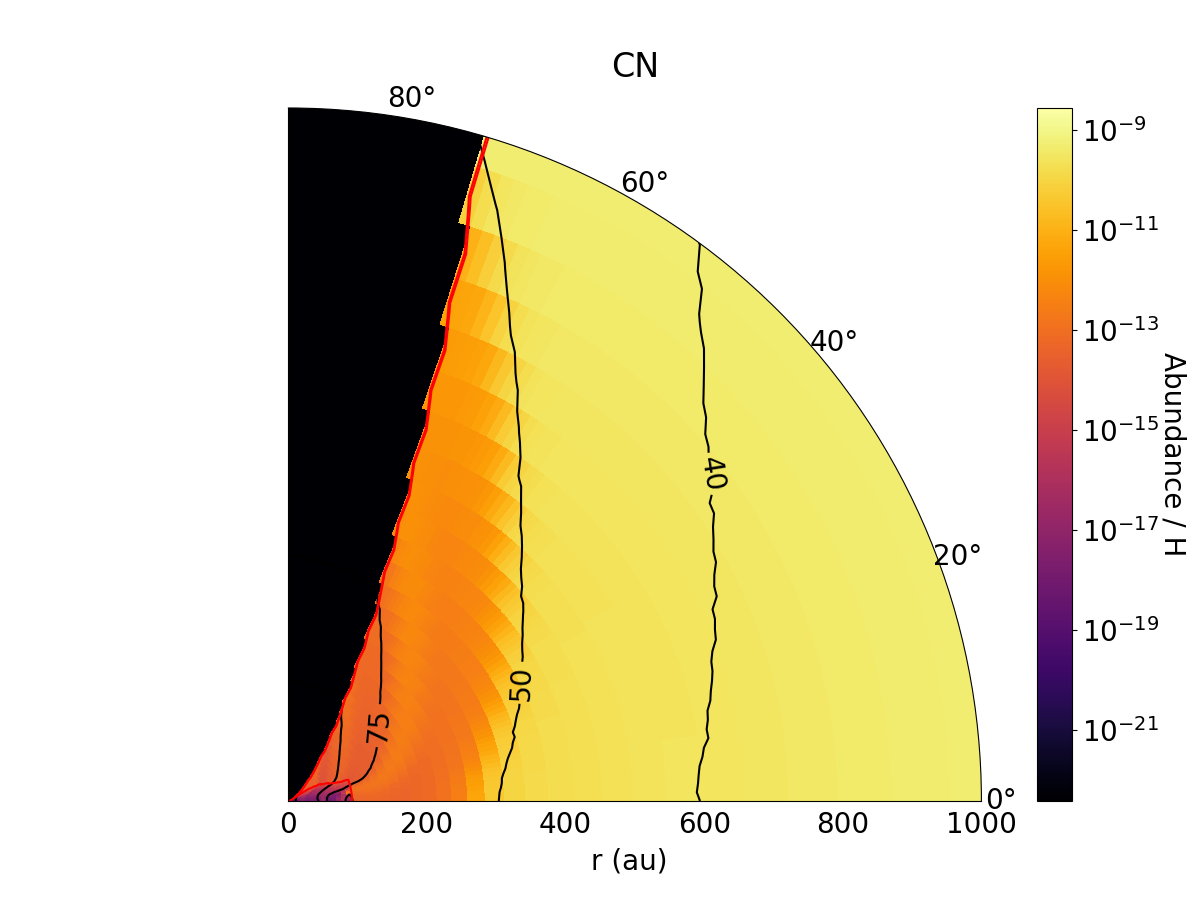}
    \includegraphics[width=0.32\textwidth, trim=6cm 0cm 0cm 0cm,clip]{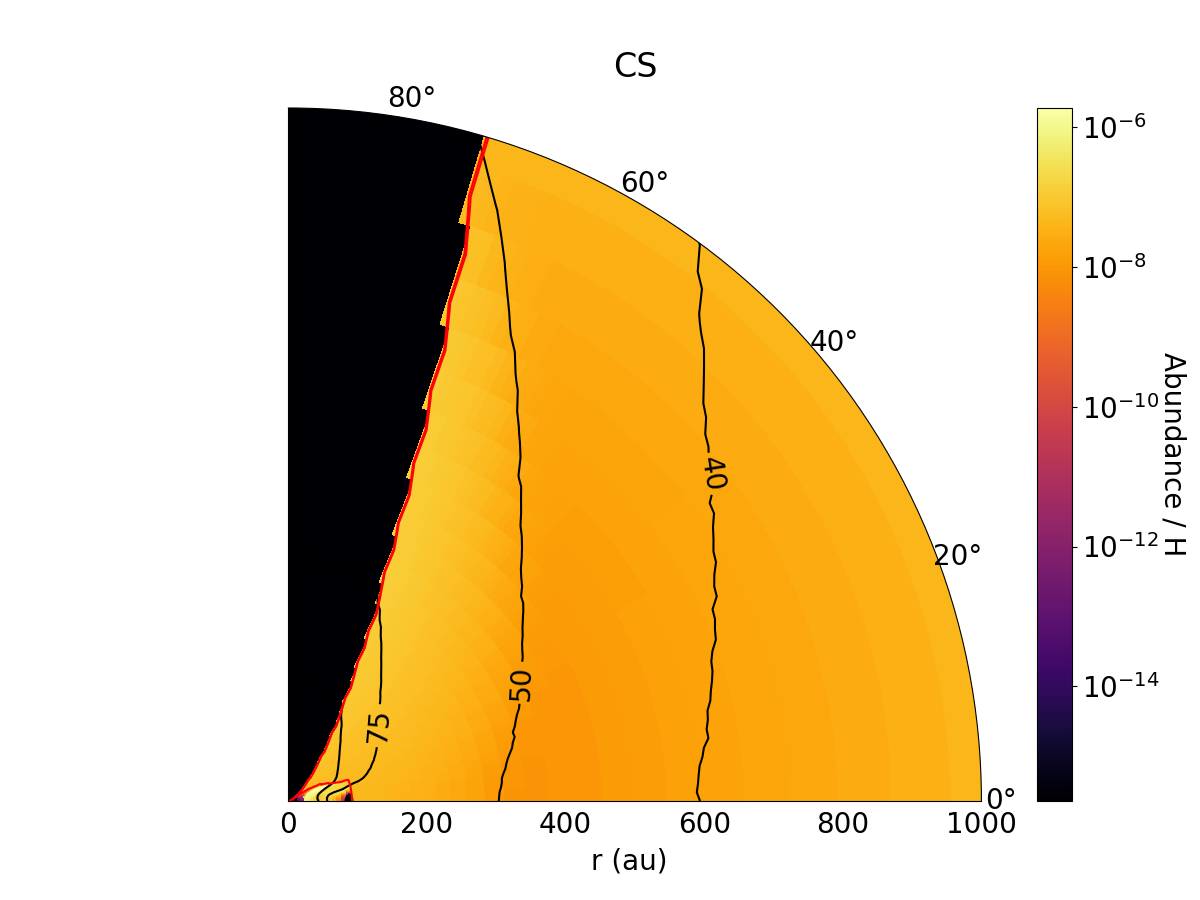}
    \includegraphics[width=0.32\textwidth, trim=6cm 0cm 0cm 0cm,clip]{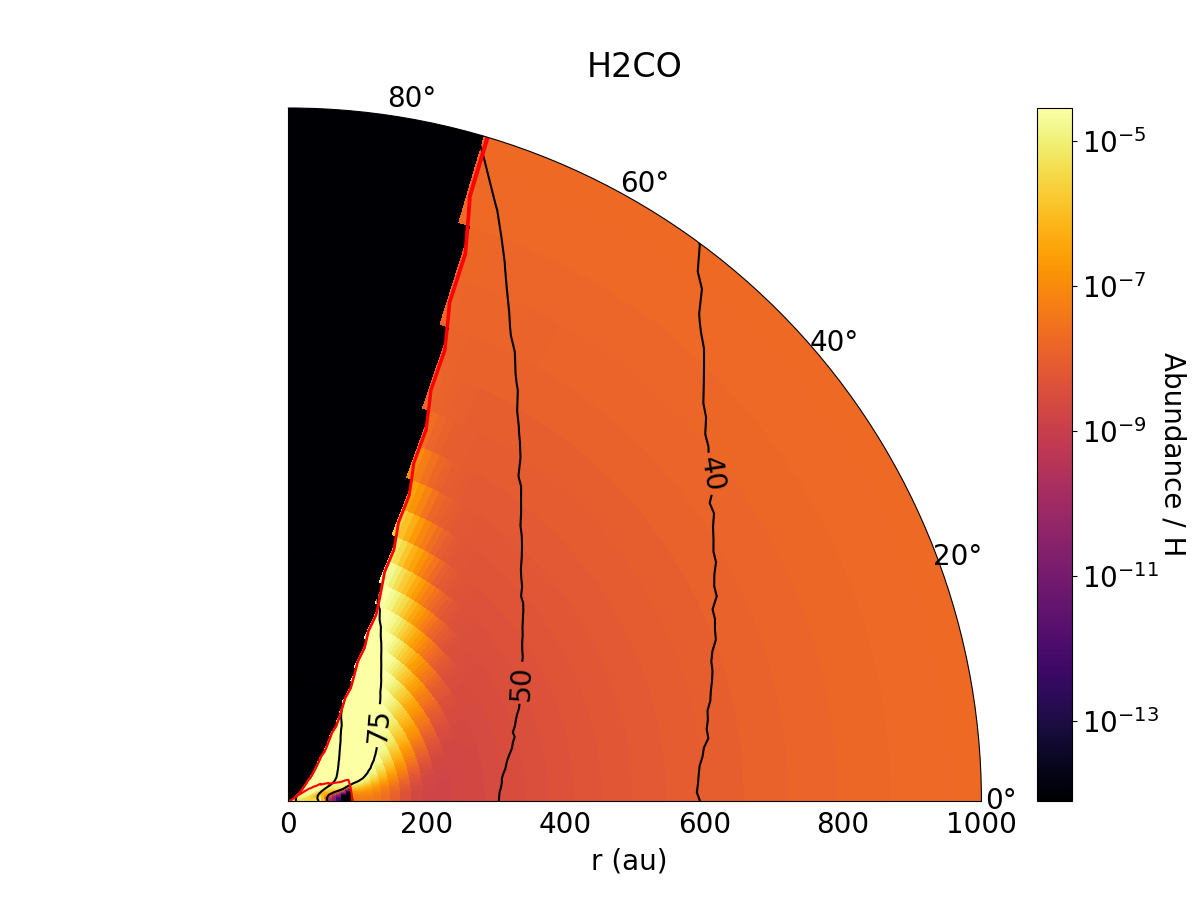}
    \includegraphics[width=0.32\textwidth, trim=6cm 0cm 0cm 0cm,clip]{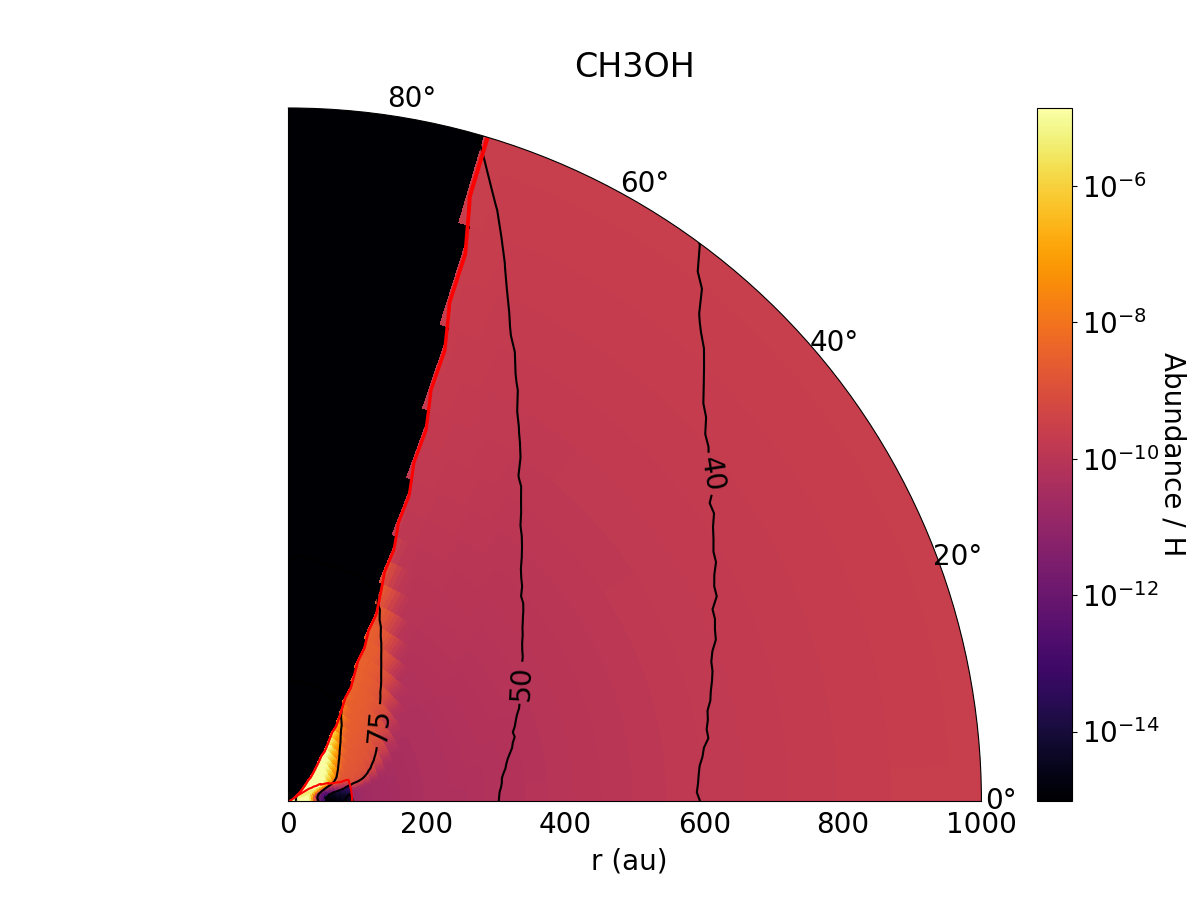}
    \caption{Reconstructed maps of relative gas phase abundances of CO (top left), CN (top middle), CS (top right), H$_2$CO (bottom left), and CH$_3$OH (bottom right), 150 kyr after the formation of the central object. The black contours represent temperatures of 40, 50, 75, 100 and 300 K. The red contours delimitate the envelope, the disk and the outflow.}
    \label{fig:maps}
\end{figure*}

We performed similar calculations for all particles in the simulation box, and use the final abundance of each particle to reconstruct abundance maps. We display resulting abundance maps at $t=\tff+150$~kyr in Fig. \ref{fig:maps}, for those five species in the gas phase.
The abundance of CO varies negligibly across the envelope. The whole computed map has a temperature larger than 30 K, which is above its desorption temperature. Conversely, CH$_3$OH is mostly frozen on grains at temperatures below 100 K. There is a clear correlation between the regions of high methanol abundance and the temperature isocontours. This map only display the gas phase abundance of methanol; the total gas+ice abundance is roughly constant throughout the collapse. This map therefore effectively illustrates the thermal desorption of the species. H$_2$CO and CS exhibit a similar behavior, with an increased abundance along the outflow cavity. On the contrary, the abundance of CN decreases in the high-temperature regions, especially above 50 K.

\subsection{Synthetic observations}\label{sec:application_obs}

\subsubsection{Observation parameters}

In this section, we present synthetic observations of the molecular emission of the species presented in Sect. \ref{sec:application_chem}:
CO, CN, CS, H$_2$CO, and CH$_3$OH, using our simulation at $t=\tff + 150$ kyr. For qualitative comparison, we mimicked the observation parameters used by \citet{Podio2020} to observe those five molecules toward the Class I protostar IRAS~04302+2247 with ALMA. We selected spectral windows in the band 6 of ALMA that were centered on the transition frequencies listed in Table \ref{tab:lines}.
The first step is generating the emission maps using RADMC-3D. The main input parameters are the abundance maps, the inclination angle, and the spectral resolution, which we set at 141 kHz ($\sim 0.18$ \kms). The spectral window spans from -10 to 10 \kms, which results in a sampling of $\sim 110$ channels.
The emission map is then used as input for Imager to simulate ALMA observations. To match the source parameters, we assumed a 90 degrees inclination of the source (edge-on disk), a source distance of 160 pc, and a declination of +22 degrees. We simulated an observation with the C-5 configuration of ALMA with a 13 minutes integration time, resulting in a beam size of $0.32 \times 0.25~\arcsec$. The noise in synthetic observations is purely numerical. The observing time therefore only affects the uv-coverage.

\begin{table}
  \caption{Properties of the molecular lines in our synthetic observations.}
  \label{tab:lines}
\centering
\begin{tabular}{llll}
\hline\hline
  Species & Quantum numbers & Frequency (GHz)\\
\hline
 CO & 2 -- 1 & 230.5380\\
 CN & 2 -- 1, J=$\frac{5}{2}$--$\frac{3}{2}$, F=$\frac{5}{2}$--$\frac{3}{2}$ & 226.8742\\
 CN & 2 -- 1, J=$\frac{5}{2}$--$\frac{3}{2}$, F=$\frac{7}{2}$--$\frac{3}{2}$ & 226.8748\\
 CN & 2 -- 1, J=$\frac{5}{2}$--$\frac{3}{2}$, F=$\frac{3}{2}$--$\frac{1}{2}$ & 226.8759\\
 CS & 5 -- 4 & 244.9356\\
 H$_2$CO & 3$_{1,2}$ -- 2$_{1,1}$ & 225.6978\\
 CH$_3$OH & 5$_{0,4}$ -- 4$_{0,4}$ & 241.7914\\
\hline
\end{tabular} 
\end{table}

\subsubsection{Results}

\begin{figure*}
    \centering
    \includegraphics[width=0.3\textwidth, trim=0cm 0cm 0cm 0cm]{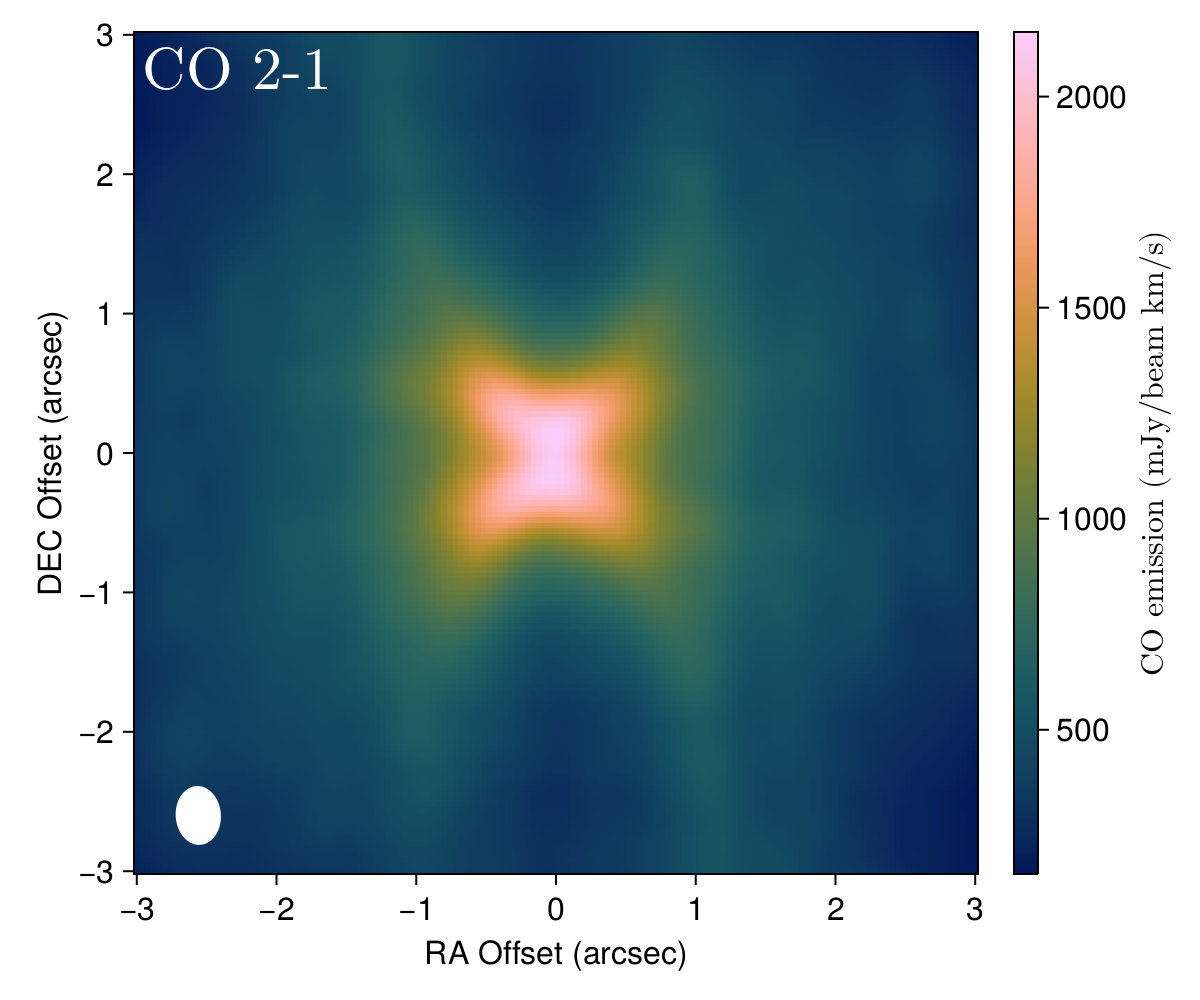}
    \includegraphics[width=0.3\textwidth, trim=0cm 0cm 0cm 0cm]{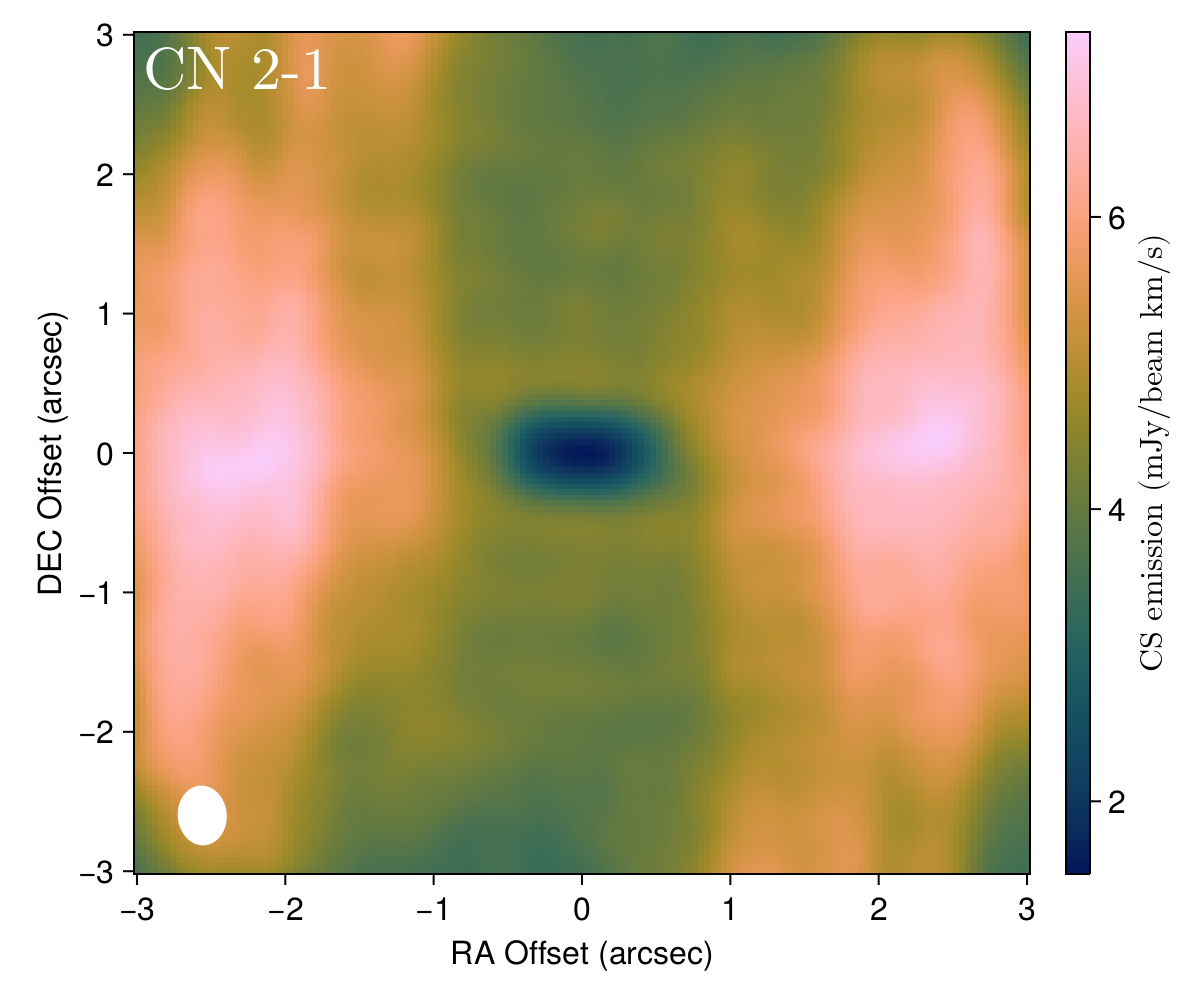}
    \includegraphics[width=0.3\textwidth, trim=0cm 0cm 0cm 0cm]{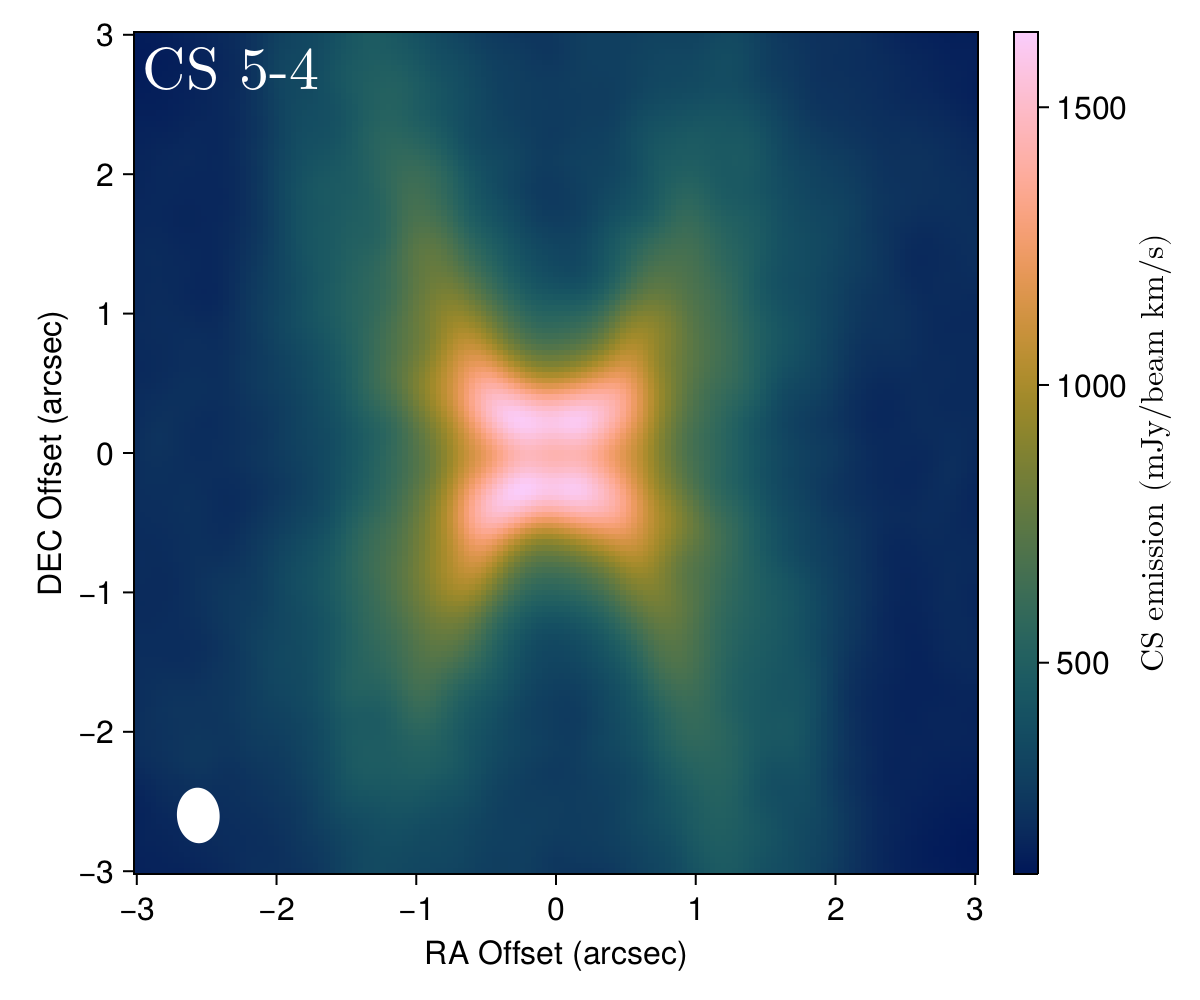}
    \includegraphics[width=0.3\textwidth, trim=0cm 0cm 0cm 0cm]{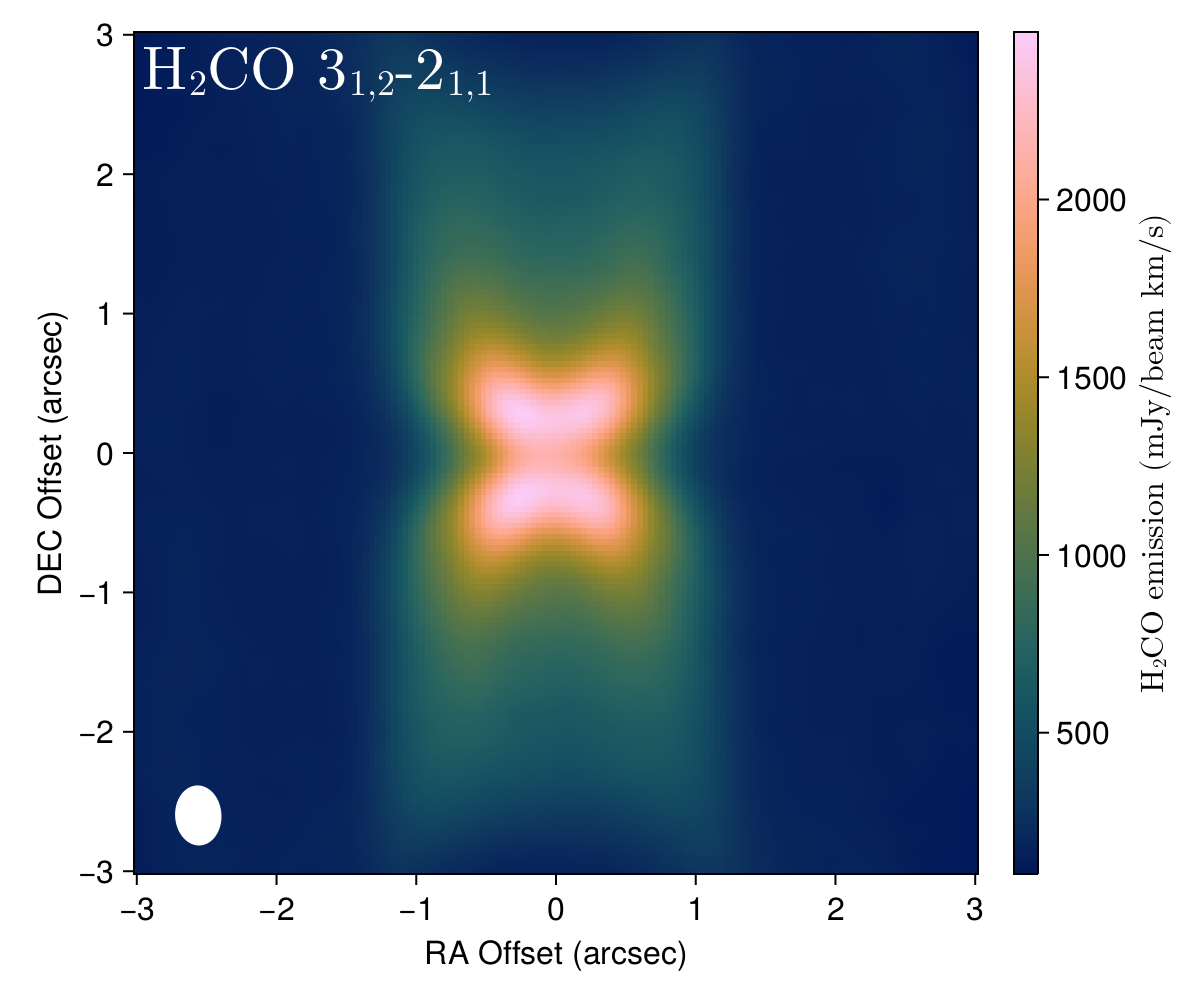}
    \includegraphics[width=0.3\textwidth, trim=0cm 0cm 0cm 0cm]{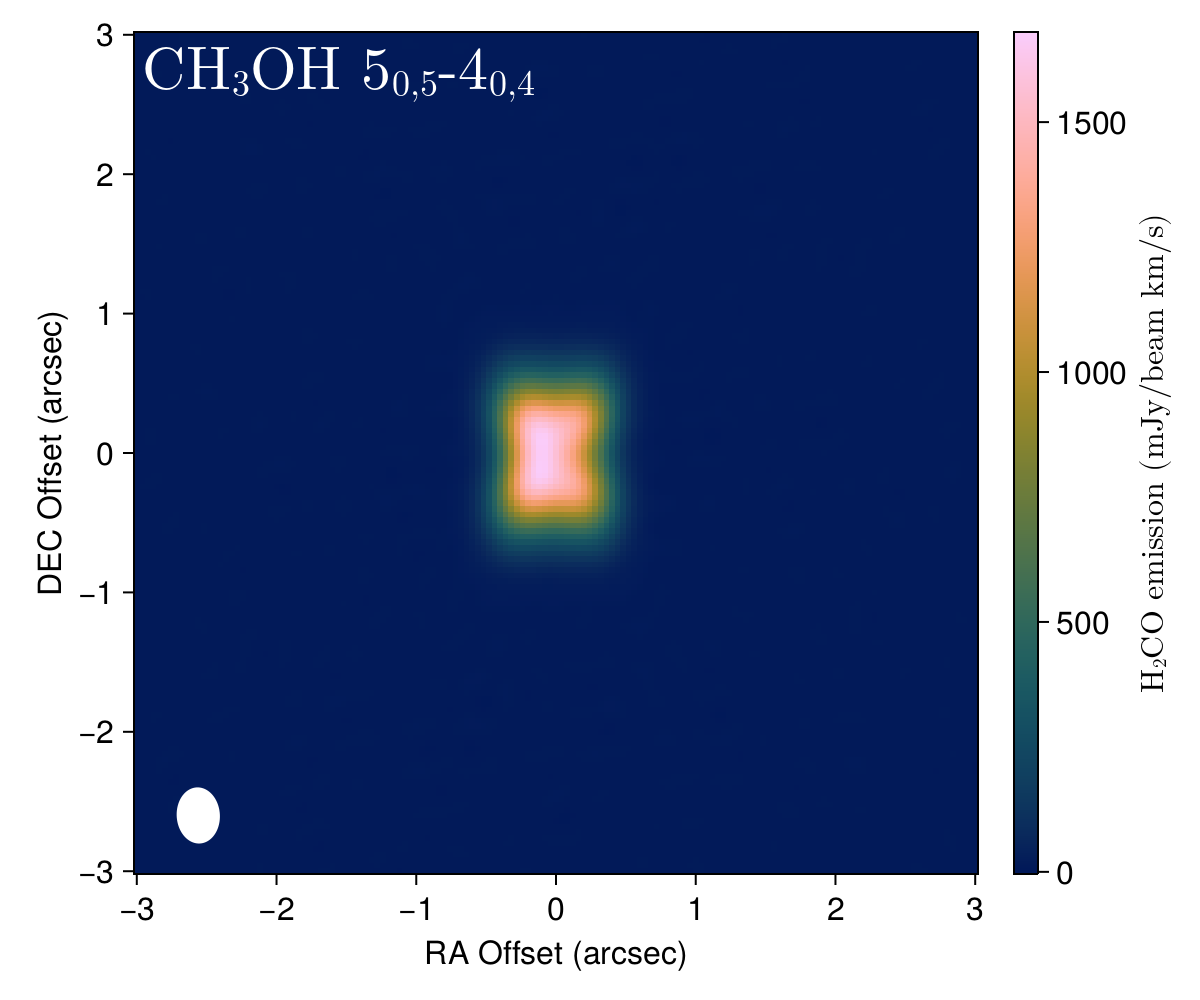}
    \caption{Moment 0 map of the molecular emission for an edge-on configuration, zoomed on the inner 6$\arcsec$. The white ellipse on the bottom left represents the beam size. From left to right, top to bottom: CO 2--1, CN 2--1, CS 5--4, H$_2$CO 3$_{1,2}$--2$_{1,1}$, CH$_3$OH 5$_{0,5}$--4$_{0,4}$}
    \label{fig:map_moment0}
\end{figure*}

We display the moment 0 map of the molecular emission in Fig. \ref{fig:map_moment0} for the inner $6\arcsec \times 6 \arcsec$ region. We see an X-shaped emission for CO, CS, and H$_2$CO. This structure corresponds to the hot regions along the cavity of the outflow, as displayed for those species in Fig \ref{fig:maps}. Those X structures are extremely similar to what is observed in IRAS~04302+2247 \citep[see Fig. 1 of][]{Podio2020}. As in \citet{Podio2020}, not much can be seen in the CN emission. However, the CH$_3$OH emission is faint in IRAS~04302+2247, while we clearly see a bright X-shape at the $\sim 1\arcsec$ scale in our simulation. The low bolometric luminosity of 0.43 L$_\odot$ measured toward IRAS~04302+2247 \citep{Ohashi2023} may not be enough to allow a significant desorption of methanol in the envelope.

\section{Conclusion} \label{sec:conclusion}

In this work, we presented the Analytical Protostellar Environment (APE) code, which we made publicly available. APE is a tool that facilitates chemical simulations and synthetic observations of protostellar systems. The code contains a physical model able to simulate the gravitational collapse of a dense core, and the formation and evolution of a central object and a protoplanetary disk throughout the Class 0 and I phases. We also provide scripts and interfaces to the publicly available codes Nautilus (for chemical simulations), RADMC-3D (for radiative transfer calculations), and Imager (for synthetic imaging).

We illustrated the capabilities of APE by computing the abundance maps of CO, CN, CS, H$_2$CO, and CH$_3$OH in a Class I protostellar system. We also performed synthetic ALMA observations of those species assuming an edge-on source. The moment 0 maps of CO, CS and H$_2$CO show an X-shaped structure extremely similar to the emission seen toward the Class I protostar IRAS~04302+2247 \citep{Podio2020}.

\section*{Data availability}

APE is publicly available on Bitbucket \footnote{\url{https://bitbucket.org/pmarchan/ape_code}}. A user manual is provided with the code to detail its usage.

\begin{acknowledgements}
    We thank the anonymous referee for their comments that helped us improve the clarity of the manuscript. We thank Takashi Hosokawa for kindly allowing us to publicly use and provide his tabulated simulation results. We also thank Ugo Lebreuilly for providing his simulation results for the dust evolution. This study is part of a project that has received funding from the European Research Council (ERC) under the European Union’s Horizon 2020 research and innovation program (Grant agreement No. 949278, Chemtrip).
\end{acknowledgements}

\bibliographystyle{aa}
\bibliography{MaBiblio}

\appendix

\section{Benchmark of the envelope model}\label{app:envelope}

Since APE serves as an alternative to MHD simulations, we compare the results obtained by both methods in this section.
In classical MHD simulations, the envelope collapses, then forms an opaque first hydrostatic core with a $\lesssim 20$~au radius \citep{Larson1969}. This core contracts adiabatically for at most a few thousand years, before collapsing into a central object. A circumstellar disk forms in and around the first core and grows in size. APE does not model the first core, nor the initial growth of the disk, and the computational cost of MHD simulations increases significantly at the formation of the central object (which is the main motivation for the creation of APE). For those reasons, we restricted our comparison to the envelope before the ignition of the central object.

We used the RAMSES code \citep{Teyssier2002} with its ideal MHD solver \citep{Fromang2006}. The initial condition is a 2~M$_\odot$ critical Bonnor-Ebert sphere, with an initial radial velocity of the gas matching that of APE
\begin{equation}
    v_r(r) = -\sqrt{\frac{GM_\mathrm{in}(r)}{2r}}.
\end{equation}
We performed a run with a moderate initial magnetic field (mass-to-flux ratio $\lambda=5$). Once the initial condition is set up, the gas is subjected to the effects of gravity, thermal pressure and the Lorentz force. The temperature is imposed at $T=10$~K and the extinction at $A_\mathrm{v}=3.5$.

\begin{figure}[h!]
    \centering
    \includegraphics[width=0.5\textwidth, trim=0cm 0cm 0cm 0cm]{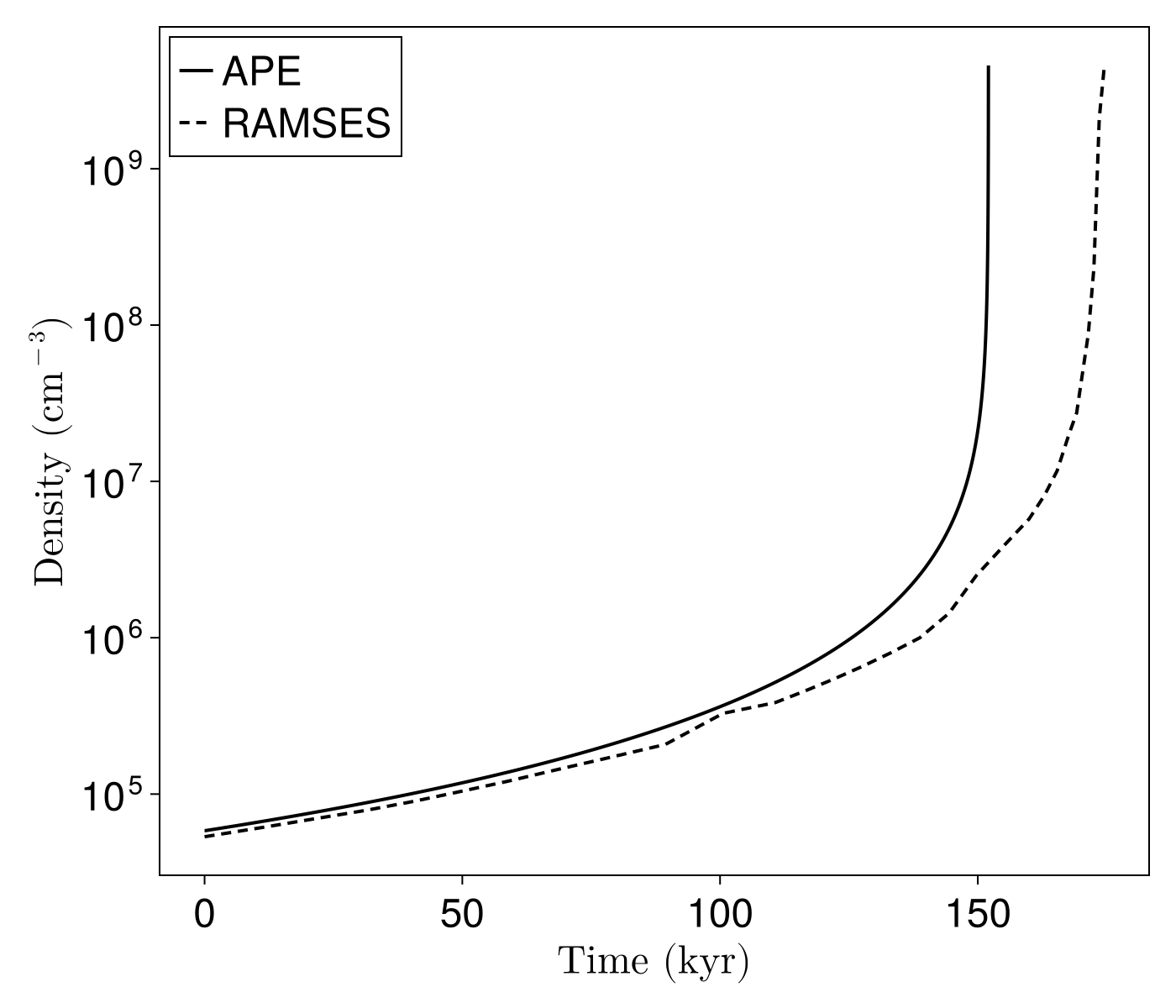}
    \caption{Time evolution of the density for a particle ending at 50~au just before the ignition of the central object, for the APE (solid line) and the RAMSES (dashed line) simulations.}
    \label{fig:app:density}
\end{figure}

We compared the evolution of particles ending at 50~au from the center in the midplane right before the ignition of the central object. The density history of the two particles is displayed in Fig. \ref{fig:app:density}. The initial $\sim 100$~kyr are nearly identical. The formation of a central object is however $\sim 15$\% longer in the MHD simulation, which results in a slightly lower density after 100~kyr. Both particles then reach a density of $\sim 4 \times 10^{9}$~cm$^{-3}$, at a time of $\sim 152$~kyr for the APE simulation, and $\sim 175$~kyr for the RAMSES simulation.

\begin{figure}[h!]
    \centering
    \includegraphics[width=0.5\textwidth, trim=0cm 0cm 0cm 0cm]{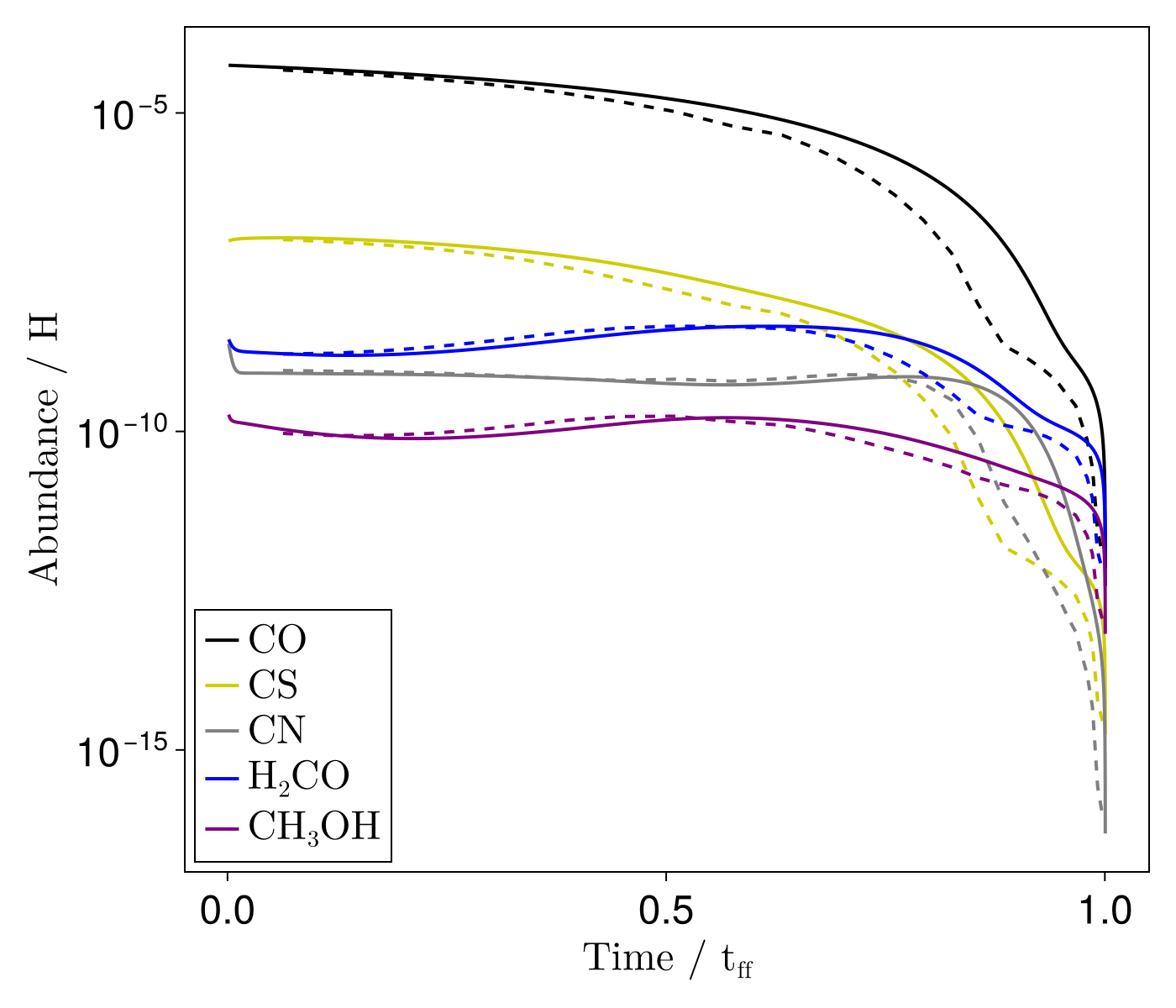}
    \caption{Time evolution of the gas-phase abundance of CO, CN, CS, H$_2$CO, and CH$_3$OH for APE (solid lines) and RAMSES (dashed lines). The time has been normalized with the formation time of the central object in each simulation.}
    \label{fig:app:abundances}
\end{figure}

We then ran Nautilus on the two particles, using the same chemical network and initial abundances as described in Sect. \ref{sec:chem_model}. As example, Fig. \ref{fig:app:abundances} displays the abundance evolution of the same five species considered in Sect. \ref{sec:application_obs}: CO, CN, CS, H$_2$CO, and CH$_3$OH. The time axis is normalized by the formation time of the central object in each simulation. Similarly to the density, the abundances are nearly identical in both simulations for the first $\sim 100 $ kyr ($t/\tff \approx 0.6$). As the abundances decrease, differences between the two cases increase up to an order of magnitude for the simplest species (CO, CN, and CS), but are less pronounced for the two more complex species (H$_2$CO and CH$_3$OH) with differences less than a factor of 2.

In the following, we considered the full chemical network. At the final time-step, all species with non-negligible abundances ($> 10^{-10}$) show differences lower than 50\% between the two simulations (with 230 out of 240 being below a 10\% difference). The only exception is gas-phase H, which is 10 times more abundant with APE ($\sim 10^{-9}$) than with RAMSES ($\sim 10^{-10}$).
Including species with abundances above $10^{-15}$, 90\% display differences smaller than 25\%, and all remain within an order of magnitude.

Overall, APE successfully reproduces the chemical trends obtained from a MHD simulation, particularly for the most abundant species, whose abundances show strong convergence. This demonstrates that APE provides a reliable alternative for studying the chemical evolution of star-forming systems, at a fraction of the cost of full MHD simulations.

\end{document}